\documentclass[journal,12pt,onecolumn,draftclsnofoot]{IEEEtran}

\IEEEoverridecommandlockouts

\usepackage{cite,cuted}
\usepackage{widetext}
\usepackage{amsmath,amssymb,amsfonts,float,graphicx,steinmetz,bm,subfigure,physics,mathtools,url,color,xcolor,etoolbox,ifthen,cite}

\newcommand{\bi}{\begin{itemize}}
\newcommand{\ei}{\end{itemize}}

\newcommand{\be}{\begin{enumerate}}
\newcommand{\ee}{\end{enumerate}}
\newcommand{\bd}{\begin{description}}
\newcommand{\ed}{\end{description}}
\newcommand{\bc}{\begin{center}}
\newcommand{\ec}{\end{center}}
\newcommand{\bt}{\begin{tabbing}}
\newcommand{\et}{\end{tabbing}}
\newcommand{\bfig}{\begin{figure}}
\newcommand{\efig}{\end{figure}}
\newcommand{\beq}{\begin{equation}}
\newcommand{\beqarr}{\begin{eqnarray}}
\newcommand{\beqarrn}{\begin{eqnarray*}}
\newcommand{\eeq}{\end{equation}}
\newcommand{\eeqarr}{\end{eqnarray}}
\newcommand{\eeqarrn}{\end{eqnarray*}}
\newcommand{\bflr}{\begin{flushright}\vspace{-0.2in}}
\newcommand{\eflr}{\end{flushright}}
\newcommand{\bsub}{\begin{subequations}}
\newcommand{\esub}{\end{subequations}}
\newcommand{\barr}{\begin{array}}
\newcommand{\earr}{\end{array}}
\newcommand{\nn}{\nonumber}

\def\undb#1{\mbox{\bf{#1}}}

\def\dsp{\displaystyle}

\def\dn{\stackrel{\scriptscriptstyle \triangle}{=}}

\begin{document}
\title{
Channel Estimation and Secret Key Rate Analysis of MIMO Terahertz Quantum Key Distribution
\thanks{ \textbf{This work has been submitted to the IEEE for possible publication. Copyright may be transferred without notice, after which this version may no longer be accessible.}
The work of Neel Kanth Kundu and Matthew R. McKay was supported by the Hong Kong Research Grants Council (grant
number C6012-20G). The work of Soumya P. Dash was supported by the
Science and Engineering Research Board (SERB), a Statutory Body of
the Department of Science and Technology (DST), Government of
India, through its Start-up Research Grant (SRG) under Grant
SRG/2019/001234. The work of Ranjan K. Mallik was supported in
part by the SERB, a Statutory Body of the DST, Government of
India, under the J. C. Bose Fellowship.
}
}

\author{
Neel Kanth Kundu, {\em Graduate Student Member, IEEE}, Soumya P. Dash, {\em Member, IEEE},
Matthew R. McKay, {\em Fellow, IEEE}, and Ranjan K. Mallik, {\em Fellow, IEEE}
\thanks{
N. K. Kundu and M. R. McKay are with the Department of
Electronic and Computer Engineering, The Hong Kong University of
Science and Technology, Clear Water Bay, Kowloon, Hong Kong
(e-mail: nkkundu@connect.ust.hk,  m.mckay@ust.hk).

S. P. Dash is with the School of Electrical Sciences, Indian Institute of Technology Bhubaneswar, Odisha, India (e-mail:  soumyapdashiitbbs@gmail.com).

R. K. Mallik is with the Department of Electrical Engineering, Indian Institute of Technology Delhi, New Delhi, India (e-mail: rkmallik@ee.iitd.ernet.in). }
\vspace{-1cm}
}
\markboth{This work has been submitted to the IEEE for possible publication. Copyright may be transferred without notice.}{This work has been submitted to the IEEE for possible publication. Copyright may be transferred without notice.}
\maketitle
\begin{abstract}
We study the secret key rate (SKR) of a multiple-input
multiple-output (MIMO) continuous variable quantum key
distribution (CVQKD) system operating at terahertz (THz)
frequencies, accounting for the effects of channel estimation. We
propose a practical channel estimation scheme for the THz MIMO
CVQKD system which is necessary to realize transmit-receive
beamforming between Alice and Bob. We characterize the
input-output relation between Alice and Bob during the key
generation phase, by incorporating the effects of additional noise
terms arising due to the channel estimation error and detector
noise. Furthermore, we analyze the SKR of the system and study the
effect of channel estimation error and overhead. Our simulation
results reveal that the SKR may degrade significantly as compared
to the SKR upper bound
that assumes
perfect channel state information, particularly at large
transmission distances.
\end{abstract}


\newpage

\IEEEpeerreviewmaketitle

\section{Introduction}
With the widespread deployment of fifth-generation (5G) wireless
communication systems, researchers have started to conceptualize
new use cases and the required technological solutions for beyond
fifth generation (B5G) or sixth generation (6G) communication
systems \cite{yang20196g}. The future B5G/6G networks aim to
support a peak data rate of $1$ Tbps, an air latency of $0.1$ ms,
and twice the spectral and energy efficiency of current 5G
standards
\cite{dang2020should,yang20196g,david20186g,giordani2020toward,rappaport2019wireless}. Different physical layer solutions have been proposed to meet the demands of B5G wireless applications spanning holographic telepresence, tactile internet, internet of everything, and augmented and virtual reality \cite{giordani2020toward}. These include multiple-input multiple-output (MIMO) systems \cite{faisal2020ultramassive}, reconfigurable
intelligent surfaces \cite{pan2021reconfigurable,di2020smart,huang2020holographic,kundu2020ris,kundu2021channel,kundu2021large}, novel modulation schemes\cite{basar2020reconfigurable,tusha2020multidimensional,zhong2018spatial,kundu2020signal},
and harnessing of the terahertz (THz) frequency spectrum \cite{sarieddeen2020next,akyildiz2014terahertz,kurner2014towards,huq2019terahertz,busari2019terahertz}.


Apart from high data-rate requirements, security and privacy of
the data are also considered to be of great importance in B5G
applications. With the rapid advancement in quantum computing,
standard higher layer encryption schemes based on the
Rivest-Shamir-Adleman (RSA) algorithm can be broken by Shor's
factoring algorithm
\cite{manzalini2020quantum,weedbrook2012gaussian}.
Similarly, physical layer encryption based on classical key
distribution algorithms like Diffie-Hellman \cite{diffie1976new} are also not secure, since its security is based on the assumption that the computationally hard problem of discrete logarithm cannot be solved in reasonable time by classical computers. Hence, current computationally secure encryption algorithms can be broken with the rapid development in practical quantum computing.
Quantum key
distribution (QKD) can be used to distribute secure keys between
two parties, say Alice and Bob, which can then be used for
one-time-pad (OTP) based physical layer encryption for 6G
applications
\cite{sanenga2020overview,al2021use,wang2020security,wang2021quantum}.
Alternatively, the key generated from a QKD protocol can be used
by the higher layers for symmetric key encryption. QKD offers
unconditional security guaranteed by the laws of quantum physics.

Broadly speaking, there are two main classes of QKD which have
been proposed in the literature. The first is discrete variable
QKD (DVQKD) that encodes the key information in the polarization
or the phase of single photon light pulses, whose security is
guaranteed by the no-cloning theorem of quantum physics
\cite{bennett1984update,bennett1992quantum,bennett1992communication,ekert1991quantum,franson1991two,inoue2002differential,buttler2002new,stucki2005fast}.
The second one is continuous variable QKD (CVQKD) that encodes the
key information in the quadratures of Gaussian coherent states,
and its security is based on the Heisenberg's uncertainty
principle
\cite{ralph2000security,hillery2000quantum,cerf2001quantum,grosshans2002continuous,grosshans2003quantum,silberhorn2002continuous}. The implementation of DVQKD is difficult in practice since it requires single photon sources and detectors. On the other hand, it is relatively easy to implement CVQKD since it requires standard
off-the-shelf telecommunication equipment. Thus, it is easier to integrate CVQKD into future wireless communication networks.


Most current wireless QKD systems are point-to-point links (e.g.,
satellite to earth links and inter-building links) implemented by
using optical frequencies
\cite{trinh2018quantum,trinh2018design,qu2017high,erven2008entangled,pirandola2020advances,pirandola2021limits,pirandola2021satellite}.
This requires high precision tracking of the receiver and does not
support mobility required for terrestrial B5G applications.
Therefore, THz QKD systems have recently been proposed for mobile
devices
\cite{ottaviani2020terahertz,busari2019terahertz,huq2019terahertz,zavitsanos2020qkd,liu2018practical},
since THz offers numerous advantages over optical frequencies such
as less delicate pointing, acquisition and tracking, and being
less affected by ambient light, atmospheric turbulence,
scintillation, cloud, and dust
\cite{elayan2019terahertz,sarieddeen2020next,akyildiz2014terahertz,rappaport2019wireless,kurner2014towards}.
Microwave frequency is not a feasible frequency spectrum for QKD, since the preparation vacuum thermal noise is much larger at room temperature at lower frequency spectrum. Therefore, THz frequency is a potential frequency spectrum for QKD applications since positive secret
key rate (SKR) is achievable at room temperature due to lower thermal noise at THz frequencies
\cite{kundu2021mimo,ottaviani2020terahertz,liu2018practical}.

Some recent studies have investigated the viability of THz CVQKD for both terrestrial
\cite{kundu2021mimo,ottaviani2020terahertz,he2020indoor,liu2018practical,liu2021multicarrier}, and inter-satellite links \cite{wang2019inter,liu2021multicarrier}. One limiting factor of THz QKD is the low SKR and maximum transmission distance due to the high free-space path loss and atmospheric absorption loss at THz frequency spectrum.
We recently proposed a MIMO THz CVQKD system that achieves a high SKR and large transmission distances by using multiple transmit and receive antennas \cite{kundu2021mimo}.
Our initial work demonstrated the feasibility of MIMO transmission
for CVQKD applications, assuming the availability of perfect
channel knowledge at Alice and Bob. However, in practice, the MIMO
channel needs to be estimated and the effect of channel estimation
error should be incorporated in the SKR analysis of the system.
This motivates us for the current work where we propose a
practical channel estimation protocol for the MIMO THz CVQKD
system, and incorporate the effect of channel estimation errors in
the input-output model during the key generation phase.
The main contributions of the paper
can be summarized as follows:
\begin{itemize}
\item  We propose a practical channel estimation protocol for the MIMO THz CVQKD system of \cite{kundu2021mimo}.
\item  We characterize the input-output relation between Alice and
Bob during the key generation phase by incorporating the
additional noise terms arising due to channel estimation errors
and detector noise. \item In contrast to our previous work
\cite{kundu2021mimo}, where we considered only the Gaussian
collective attack, here we consider two types of attacks implementable by Eve: individual and collective attacks. In the former case, the maximum key information that Eve can steal is given by Shannon's mutual information while in the latter stronger attack the maximum key information that Eve can steal is bounded by the Holevo information between Bob's output state and Eve's ancilla state. The type of attack that Eve can implement depends on the quantum resources available to her.


\item We analyze the SKR for both types of attacks by
incorporating the effects of channel estimation overhead, the
additional noise terms due to imperfect channel estimation, and
the detector noise at Bob.
\item We study the effect of channel estimation error on the SKR
of the MIMO CVQKD system, and analyze the effect of key parameters
such as pilot length and pilot power on the SKR of the MIMO CVQKD
system. We also study the maximum threshold on the noise variance
(arising due to channel estimation error) that the MIMO CVQKD
system can tolerate in order to attain positive SKRs.
\end{itemize}

The rest of the paper is organized as follows. Section
\ref{system} presents the system model, the channel estimation
protocol, and the input-output relation between Alice and Bob
obtained from SVD based transmit-receive beamforming with
imperfect channel state information. The SKR analysis for both the
individual and collective attacks are presented in Section
\ref{SKR}. Simulation results are shown in Section
\ref{simulation} and finally some concluding remarks are made in
Section \ref{conclusion}.

{\em Notation}: Boldface ($\undb{A}$)
letters are used for representing matrices. $\undb{A}^{\dagger}$ and $\undb{A}^{T}$ denote the conjugate transpose and transpose of a matrix $\undb{A}$, respectively. A matrix of all ones and all
zeros is represented by $\bm{1}_{M\times N} \,, \bm{0}_{M\times N} \in
{\mathbb C}^{M\times N}$, respectively, an $M\times M$ identity matrix is denoted by $\bm{I}_M$, and ${\rm diag}(\bm{a})$ with $\bm{a} \in {\mathbb C}^{M}$
returns an $M\times M$ diagonal matrix with the elements of
$\bm{a}$ on its diagonals. A real multivariate Gaussian distribution with
mean vector $\bm{\mu} \in \mathbb{R}^N$ and covariance matrix $\bm{\Sigma} \in \mathbb{R}^{N\times N}$ is denoted by $\mathcal{N}\left(\bm{\mu},\bm{\Sigma}
\right)$, and a multivariate complex Gaussian distribution is denoted by $\mathcal{CN}\left(\bm{\theta},\bm{\Gamma}
\right)$ where $\bm{\theta} \in \mathbb{C}^N $ is the mean vector and $\bm{\Gamma} \in \mathbb{C}^{N\times N}$ is the covariance matrix. Finally, $\text{det}(\undb{A})$ denotes the determinant of the square matrix $\undb{A}$.

\section{System Model} \label{system}
\subsection{Channel Model}
We consider two communicating parties Alice and Bob each having multiple antennas, who wish to share a quantum secure key. We assume that Alice and Bob have $N_t$ and $N_r$ antennas, respectively. The MIMO channel
$\undb{H} \in \mathbb{C}^{N_r \times N_t}$
between Alice and Bob can be modeled as
\cite{busari2019terahertz,deng2014mm}
\begin{equation}
  \undb{H} = \sum_{l=1}^{L} \sqrt{\gamma_l}  e^{j2\pi f_c \tau_l} \bm{\psi}_R\left(\phi_l^r\right) \bm{\psi}_T^{\dagger}\left(\phi_l^t\right) \;,
  \label{sys1}
\end{equation}
where $f_c$ and $L$ denote the frequency of the carrier signal and total number of multipaths, respectively. Furthermore, $\gamma_l$ and $\tau_l$  denote the the path loss and propagation delay of the $l$-th multipath, respectively. Moreover, $\phi_l^r$ denotes the angle of arrival for Bob's uniform linear array (ULA) at its $l$-th multipath component, and $\phi_l^t$ denotes the angle of departure from Alice's ULA at its $l$-th multipath component.
For the ULAs, the array response
vectors $\bm{\psi}_R\left(\phi_l^r\right)$ and
$\bm{\psi}_T\left(\phi_l^t\right) $ are given by
\begin{align}
 \bm{\psi}_R\left(\phi_l^r\right) &= \frac{1}{\sqrt{N_r}} [1, e^{j\frac{2\pi}{\lambda}d_r\sin\phi_l^r},\ldots, e^{j\frac{2\pi}{\lambda}d_r(N_r-1)\sin\phi_l^r}]^T \, ,\nonumber \\
  \bm{\psi}_T\left(\phi_l^t\right) &= \frac{1}{\sqrt{N_t}} [1, e^{j\frac{2\pi}{\lambda}d_t\sin\phi_l^t},\ldots, e^{j\frac{2\pi}{\lambda}d_t(N_t-1)\sin\phi_l^t}]^T \, ,
 \label{sys2}
\end{align}
where $d_t,d_r $ are the inter-antenna spacings at Alice's and
Bob's ULAs, respectively, and $\lambda$ denotes the wavelength of the carrier signal. 
In the channel model (\ref{sys1}), $\gamma_l$ denotes the path loss which can be modelled as
\cite{he2020indoor}
\begin{eqnarray}
  \gamma_l = \left\{
    \begin{array}{@{}ll@{}}
    \dsp \left( \frac{\lambda}{4\pi d_l}\right)^2 G_t G_r 10^{-0.1\delta d_l} , & \! \!
    l = 1  \; \textnormal{(LoS)}\; , \\
     \dsp \beta r_l \left( \frac{\lambda}{4\pi d_l}\right)^2 G_t G_r 10^{-0.1\delta d_l} , & \! \! l=2,3,\ldots,L \; \textnormal{(NLoS)}\; ,
  \end{array} \right.
   \label{sys_los_nlos}
\end{eqnarray}
where LoS and NLoS denote line-of-sight and non-line-of-sight path, respectively, $d_l$ denotes the corresponding path length, and $\delta$ denotes the atmospheric absorption coefficient in dB/km. Furthermore, $\beta$ denotes the Rayleigh roughness factor of the scattering objects, $r_l$ denotes the Fresnel reflection coefficient of the surface encountered by the $l$-th multipath component. The array gains of Bob's and Alice's ULAs are denoted by $G_r$ and $G_t$, respectively which depend on the antennas gain of each element $G_a$ as \cite{sun2018propagation}
\begin{align}
  G_r = N_r G_a \,, \; G_t = N_t G_a \;.
  \label{sys_G}
\end{align}
Similar to our initial work on THz MIMO CVQKD \cite{kundu2021mimo}, {\em we
incorporate the effects of both free-space path loss along with the atmospheric attenuation loss}, in contrast to the earlier works on THz CVQKD \cite{ottaviani2020terahertz,liu2018practical} which did not consider the free-space path loss component in the channel model.

\subsection{Channel Estimation}
\begin{figure}[ht] 
\centering
\includegraphics[width=0.8\textwidth]{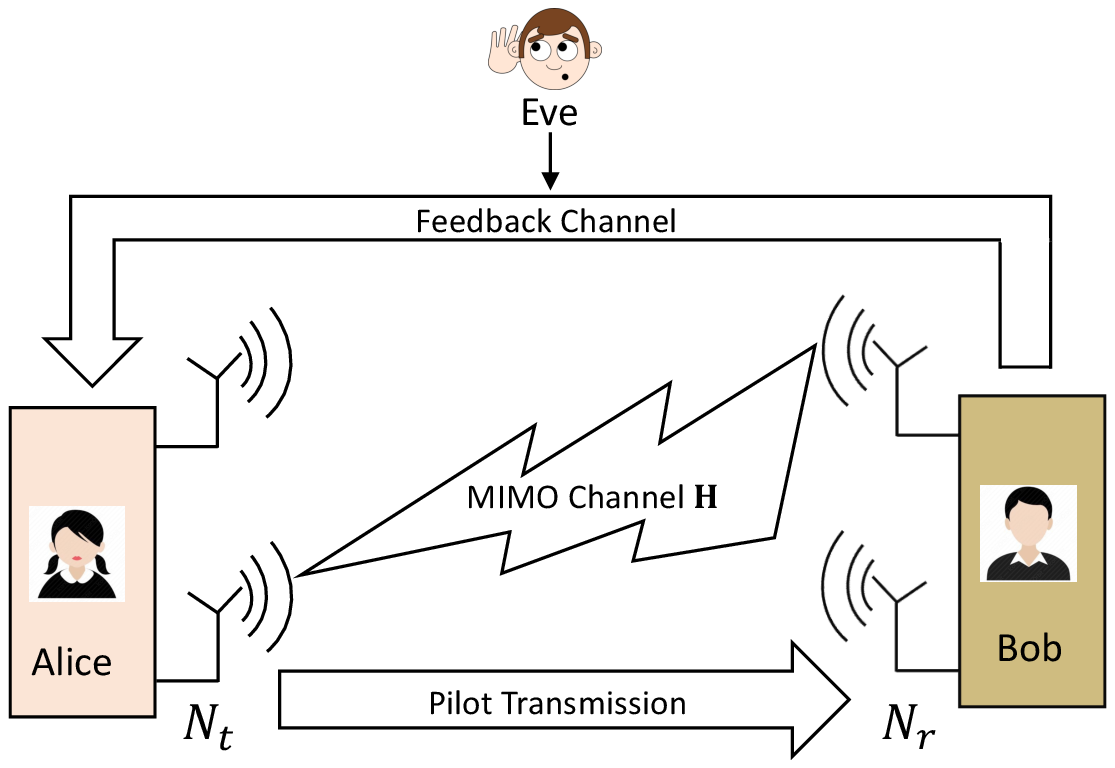}%
\caption{A schematic diagram of channel estimation protocol in
which Eve gains the channel knowledge by intercepting the
classical feedback channel.} \label{fig1_ch}
\end{figure}

We consider a MIMO CVQKD system where the wireless channel between Alice and Bob is estimated by Bob prior to the deployment of the actual key distribution protocol. We assume a perfect feedback link between Bob and Alice such that the estimated channel parameters are fed back to Alice by Bob via a public authenticated channel. Furthermore, we consider that Eve does not have the knowledge of the wireless channel initially, and she tries to gain knowledge of the MIMO channel matrix by intercepting the feedback link. Additionally, we assume that the best channel estimate that Eve can attain is the channel estimated by Bob during the channel estimation phase. A schematic diagram of the channel estimation protocol with the classical feedback channel is shown in Fig. \ref{fig1_ch}.

During the $t$-th pilot transmission phase, Alice prepares $N_t$ Gaussian coherent states $\ket{\alpha_{p,i}^t}$ with $\alpha_{p,i}^t =
q_{p,i}^t+jp_{p,i}^t \,,\, \forall \, i=1,\ldots,N_t$, which are then transmitted them from the $N_t$ antennas. The signal power during
the pilot transmission phase is $V_p$ such that
$\mathbb{E}[(q_{p,i}^t)^2]=\mathbb{E}[(p_{p,i}^t)^2]=V_p$, with
$\mathbb{E}[\cdot]$ denoting the expectation operator. The
transmitted pilot signal modes from Alice during the $t$-th pilot
transmission phase is denoted as
$\undb{x}_p^t=\undb{q}_{p}^t+j\undb{p}_{p}^t$, where
$\undb{q}_{p}^t = [q_{p,1}^t,\ldots, q_{p,N_t}^t ]^T $ and
$\undb{p}_{p}^t = [p_{p,1}^t,\ldots, p_{p,N_t}^t ]^T $.
%
After receiving the signal modes, Bob performs heterodyne
measurement to measure both quadratures of the received mode. This
results in the following input-output relation for the $t$-th
pilot transmission phase given by
\begin{align}
     \undb{y}^t = \undb{H} \undb{x}_p^t + \undb{H} \undb{x}_0^t +  \undb{n}_{{\rm het}}^t \, ,
     \label{ch2}
\end{align}
where ${\rm Re}\{\undb{y}^t\} = \hat{\undb{X}}_{B,I}^t,\; {\rm
Im}\{\undb{y}^t\} = \hat{\undb{X}}_{B,Q}^t $ are the in-phase and
quadrature phase components, respectively, of the received mode at
Bob after performing the heterodyne measurement, and
$\undb{n}_{{\rm het}}^t =\undb{n}_{{\rm het},I}^t+j\undb{n}_{{\rm
het},Q}^t  $ is the additive receiver noise due to heterodyne
measurement with  $\undb{n}_{{\rm het},I}^t,\undb{n}_{{\rm
het},Q}^t \sim \mathcal{N}\left(\bm{0}_{N_r \times 1},(2v_{\rm
el}+1)\bm{I}_{N_r} \right)$, where $v_{\rm el}$ is the variance of
the electronic noise \cite{djordjevic}.  Furthermore,
$\undb{x}_0^t= \undb{q}_{0}^t+j\undb{p}_{0}^t$ is the preparation
thermal noise at Alice with $\undb{q}_0^t, \undb{p}_0^t \sim
\mathcal{N}\left(\bm{0}_{N_t \times 1},V_0\bm{I}_{N_t} \right)$.
Here $V_0$ is the thermal noise variance given by $V_0=2\bar{n}+1$
with $\bar{n} = \left[\exp(hf_c/\kappa_BT_e)-1\right]^{-1}$, where $h$ and $\kappa_B$ denote the Planck's and Boltzmann's constants, respectively and $T_e$ is the environmental temperature in Kelvin.

We assume a quasi-static channel model where the channel matrix
$\undb{H}$ remains constant over the coherence time of the channel
$T_c$. Let $T_p<T_c$ be the pilot duration. Collecting all the
received signal modes at Bob over $t=1,2,\ldots,T_p$, the
equivalent signal model can be written as
\begin{equation}
    \undb{Y}_p = \undb{H}\undb{X}_p + \undb{H}\undb{X}_0 + \undb{N}_{\rm het} \, ,
     \label{ch3}
\end{equation}
where $\undb{Y}_p=[\undb{y}^1,\ldots,\undb{y}^{T_p}] \in
\mathbb{C}^{N_r \times T_p} $ is the matrix containing the
received signals at Bob,
$\undb{X}_p=[\undb{x}_p^1,\ldots,\undb{x}_p^{T_p}]\in
\mathbb{C}^{N_t \times T_p} $ is the matrix containing the
transmitted pilot signals from Alice,
$\undb{X}_0=[\undb{x}_0^1,\ldots,\undb{x}_0^{T_p}] \in
\mathbb{C}^{N_t \times T_p} $ contains the unknown preparation
thermal noise, and $\undb{N}_{\rm het}=[\undb{n}_{\rm
het}^1,\ldots,\undb{n}_{\rm het}^{T_p}] \in \mathbb{C}^{N_r \times
T_p} $ contains the additive electronic noise at Bob. Alice and
Bob agree upon a fixed pilot matrix $\undb{X}_p$ over a classical
public channel for the purpose of channel estimation. As such,
$\undb{X}_p $ is perfectly known to both Alice and Bob. The
problem of channel estimation requires estimating the unknown
matrix $\undb{H}$ from the equivalent linear measurement model
\begin{equation}
     \undb{Y}_p = \undb{H}\undb{X}_p + \Tilde{\undb{N}} \, ,
     \label{ch4}
\end{equation}
where $\Tilde{\undb{N}}=\undb{H}\undb{X}_0 + \undb{N}_{\rm het}$
is the equivalent noise matrix. Note that the covariance matrix of
$\Tilde{\undb{N}}$ is unknown since $\undb{H}$ is unknown. To
estimate $ \undb{H}$, we employ a least squares (LS) scheme which
leads to
\begin{equation}
    \undb{H}_{\rm LS} = \undb{Y}_p\undb{X}_{p}^{+} \, ,
    \label{ch5}
\end{equation}
where $\undb{X}_{p}^{+} = \undb{X}_{p}^{\dagger}\left(\undb{X}_{p}\undb{X}_{p}^{\dagger} \right)^{-1} $.

\subsubsection{Optimal Pilot Matrix}
We now find the optimal pilot matrix $\undb{X}_{p}$ that minimizes
the channel estimation error. Substituting (\ref{ch4}) in
(\ref{ch5}), we obtain
\begin{equation}
    \undb{H}_{\rm LS} = \undb{H} + \underbrace{\tilde{\undb{N}}\undb{X}_{p}^{+}}_{\Delta\undb{H}} \;.
    \label{ch6}
\end{equation}
The optimal pilot matrix that minimizes the mean squared error
$\mathbb{E}\left[ \|\Delta \undb{H} \|_F^2 \right] =
\mathbb{E}\left[ {\rm
tr}\left(\Delta\undb{H}\Delta\undb{H}^{\dagger} \right) \right]$
can be obtained by solving the following optimization problem
\begin{equation}
    \begin{split}
    & \underset{ \undb{X}_p} {{\rm min}} \quad {\rm tr} \left( \; \mathbb{E}  \left[ \tilde{\undb{N}} \undb{X}_p^{\dagger} \left(\undb{X}_p \undb{X}_p^{\dagger}\right)^{-2}\undb{X}_p \tilde{\undb{N}}^{\dagger}  \right] \right)\\ &
    {\rm s.t} \quad \quad {\rm tr}\left(\undb{X}_{p}^{\dagger}\undb{X}_{p} \right)=  V_p N_tT_p \, .
    \end{split}
    \label{ch7}
\end{equation}
We note that the columns of the noise matrix $\tilde{\undb{N}} $
are independent and identically distributed Gaussian random
vectors. Let $\undb{C}_n$ be the covariance matrix of the columns
of $\tilde{\undb{N}}$; we then have $\tilde{\undb{N}} \sim
\mathcal{CN}_{N_r, T_p}\left(\mathbf{0}_{N_r \times T_p},
\undb{C}_n \otimes \mathbf{I}_{T_p}\right)$. Using the result from
\cite[Lemma 4]{mckay2005general} for the mean of a matrix-variate
complex quadratic form, the equivalent optimization problem is given by
\begin{equation}
    \begin{split}
    & \underset{ \undb{X}_p} {{\rm min}} \quad {\rm tr} \left(   \left(\undb{X}_p \undb{X}_p^{\dagger}\right)^{-3}   \right) {\rm tr} \left( \undb{C}_n \right)\\ &
    {\rm s.t} \quad \quad {\rm tr}\left(\undb{X}_{p}^{\dagger}\undb{X}_{p} \right)=  V_p N_tT_p \, .
    \end{split}
    \label{ch8}
\end{equation}
The optimal $ \undb{X}_p$ that minimizes the objective satisfies $\undb{X}_p \undb{X}_p^{\dagger} = (V_p T_p) \undb{I}_{N_t} $ \cite{manton2001optimal,biguesh2006training}. Thus, $\undb{X}_p $ should contain orthogonal rows with the norm of each row being equal to $\sqrt{V_p T_p}$. One particular solution is constructed from the discrete Fourier transform (DFT) matrix, given by
\begin{equation}
\undb{X}_p=\sqrt{V_p}\left[\begin{array}{cccc}
1 & 1 & \cdots & 1 \\
1 & W_{T_p} & \cdots & W_{T_p}^{T_p-1} \\
\vdots & \vdots & & \vdots \\
1 & W_{T_p}^{N_t-1} & \cdots & W_{T_p}^{(N_t-1)(T_p-1)}
\end{array}\right] \, ,
\label{ch9}
\end{equation}
where $W_{T_p} = e^{j2\pi/T_p} $. This will be applied throughout the rest of the paper.

\subsection{Key Generation}
In this subsection we characterize the input-output relation
between Alice and Bob during the key generation phase obtained
from SVD-based transmit-receive beamforming. In contrast to our
previous work \cite{kundu2021mimo} that assumed perfect channel
knowledge, here we incorporate the effects of channel estimation
error in the input-output model.

During the key generation phase, Alice employs Gaussian modulation for encoding the key information. She generates two statistically independent random vectors, $\bm{p}_A$ and $\bm{q}_A$, that follow a Gaussian distribution, i.e.,
$\bm{p}_A, \bm{q}_A\sim \mathcal{N}\left(\bm{0}_{N_t \times
1},V_s\bm{I}_{N_t} \right)$, where $V_s$ denotes the power utilized for encoding the initial key information. She then generates $N_t$ displaced Gaussian coherent states denoted as $\ket{\alpha_i}$ with $\alpha_i = q_{A,i}+jp_{A,i} \,,\, \forall \, i=1,\ldots,N_t$, and radiates them from her transmit antennas.
We assume that during the key generation phase, Eve has the knowledge of $\undb{H}_{\rm LS}$ and uses it to inject her Gaussian mode. Let $\undb{H}_{\rm LS} = \undb{U}_{\rm LS} \bm{\Sigma}_{\rm LS} \undb{V}_{\rm LS}^{\dagger}
$ be the SVD of $\undb{H}_{\rm LS} $. Analogous to
\cite{kundu2021mimo}, Alice uses $ \undb{V}_{\rm LS}$ for transmit beamforming and Bob uses $\undb{U}_{\rm LS}$ for receive combining. The effective input-output relation during the key generation phase is then given by
\begin{equation}
    \hat{\bm{a}}_B = \undb{U}_{\rm LS}^{\dagger} \undb{H} \undb{V}_{\rm LS} \hat{\bm{a}}_A + \undb{U}_{\rm LS}^{\dagger} \undb{U}_{\rm LS} \undb{S}_{\rm LS} \hat{\bm{a}}_E \, ,
    \label{kg1}
\end{equation}
where $\hat{\bm{a}}_A = [\hat{a}_{A,1},\ldots,\hat{a}_{A,N_t} ]^T $ represents the vector of transmitted mode from Alice, $\hat{\bm{a}}_B = [\hat{a}_{B,1},\ldots,\hat{a}_{B,N_r}]^T$ represents the received mode vector at Bob, and $\hat{\bm{a}}_E =
[\hat{a}_{E,1},\ldots,\hat{a}_{E,N_t} ]^T$ represents the Gaussian noise vector introduced by Eve to extract the key information. Further, $\bm{\Sigma}_{\rm LS}$
and $\undb{S}_{\rm LS}$ are diagonal matrices with entries
\begin{align}
    \bm{\Sigma}_{\rm LS} &= {\rm diag}\left\{\sqrt{\hat{T}_1},\ldots,\sqrt{\hat{T}_r}, \bm{0}_{(m-r)\times 1 } \right\} \, , \nonumber \\
   \undb{S}_{\rm LS} &= {\rm diag}\left\{\sqrt{1-\hat{T}_1},\ldots,\sqrt{1-\hat{T}_r}, \bm{1}_{(m-r)\times 1 } \right\} \, ,
   \label{kg2}
\end{align}
where $m={\rm min}(N_t,N_r)$, and $\hat{T}_1,\ldots,\hat{T}_r$ denote the $r$ non-zero eigenvalues of $\undb{H}_{\rm LS}^{\dagger} \undb{H}_{\rm LS}$. Using (\ref{ch6}) in (\ref{kg1}), the equivalent input-output model admits
\begin{equation}
    \hat{\bm{a}}_B = \bm{\Sigma} \hat{\bm{a}}_A -\underbrace{ \undb{U}^{\dagger}_{\rm LS}\Delta \undb{H}\undb{V}_{\rm LS} \hat{\bm{a}}_A}_{\undb{n}_h}+  \undb{S}_{\rm LS} \hat{\bm{a}}_E \, ,
     \label{kg3}
\end{equation}
where $\undb{n}_h$ represents the additional noise term arising due to channel estimation error.

Bob performs measurement on the received signal mode in order to extract the secret key information. Note that during the channel estimation phase, Bob performs heterodyne measurement since both quadratures of the received signal should be measured in order to estimate the complex valued channel matrix $\undb{H}$. On the other hand, during the key generation phase, Bob can perform either homodyne or heterodyne measurement since the secret key can be extracted from the real-valued measurement outcome of one of the quadratures or both. Upon performing the measurement, the input-output relation between Alice and Bob in terms of the quadratures is given by
\begin{align}
 \hat{X}_{B,i}&=  \sqrt{\hat{T}_i} \hat{X}_{A,i} + \sqrt{1-\hat{T}_i} \hat{X}_{E,i} - n_{h,i} + n_{{\rm det},i} \;, \quad  i=1,2,\ldots,r \, ,
  \label{kg5}
\end{align}
where $\hat{X}_{B,i} $ represents Bob's quadrature measurement outcome,
$\hat{X}_{A,i}$ represents Alice's transmitted quadrature of the $i$-th coherent state, and $\hat{X}_{E,i} $ denotes the Gaussian noise's quadrature injected by Eve to extract the key information. Here $\hat{X}$ denotes one of the two quadratures $\{ \hat{q}, \hat{p}\}$, i.e., $\hat{X}_{A,i}= \{
\hat{q}_{A,i}, \hat{p}_{A,i}\}$, and a same notation applies for the quadratures of Bob and Eve, $\hat{X}_{B,i}$, $\hat{X}_{E,i}$. Due to the presence of preparation thermal noise of variance $V_0$, Alice's
transmitted mode has a variance of $V(\hat{X}_{A,i}) = V_a = V_s+V_0$. The Gaussian noise introduced by Eve has a power of $V(\hat{X}_{E,i} )=W$. The distribution of $n_{h,i}$ arising from channel estimation error is given by $n_{h,i} \sim \mathcal{N}( 0, \sigma_{h,i}^2 )$, with $\sigma_{h,i}^2=0.5\undb{C}_h(i,i)$, where $\undb{C}_h$ denotes the covariance matrix of the additional noise vector $\undb{n}_h$ in (\ref{kg3}). Furthermore, $n_{{\rm det},i} \sim \mathcal{N}\left(0, \sigma_{\rm
det}^2\right)$ is the detector noise with $\sigma_{\rm
det}^2=d(1+v_{el})-1$, where $d$ is the measurement parameter which takes the value $d=1$ for homodyne measurement and $d=2$ for heterodyne measurement.

\subsubsection{Estimation of Noise Covariance Matrix}
Alice and Bob estimate the SKR based on the input-output model in
(\ref{kg5}), and decide to use the secret key for encryption only
if the estimated SKR is above a threshold. In order to estimate
the SKR, Alice and Bob need to estimate the variance of the noise
terms in (\ref{kg5}). We assume that Bob's detector noise variance
$\sigma_{\rm det}^2$ is perfectly known to Bob and he only needs
to estimate $\sigma_{h,i}^2$, which depends on $\undb{C}_h$.
Therefore, in this subsection we find an estimator of
$\undb{C}_h$. Using (\ref{kg3}), the covariance matrix
$\undb{C}_h$ can be expressed as
\begin{align}
  \undb{C}_h =\mathbb{E}\left[\undb{n}_h \undb{n}_h^{\dagger} \right] &=  \mathbb{E}\left[\undb{U}_{\rm LS}^{\dagger} \tilde{\undb{N}}\undb{X}_{p}^{+}\undb{V}_{\rm LS}\hat{\bm{a}}_A \hat{\bm{a}}_{A}^{\dagger}\undb{V}_{\rm LS}^{\dagger} (\undb{X}_{p}^{+})^{\dagger} \tilde{\undb{N}}^{\dagger} \undb{U}_{\rm LS} \right] \nonumber \\
  &\stackrel{(a)}{=} 2V_a \mathbb{E}\left[\undb{U}_{\rm LS}^{\dagger} \tilde{\undb{N}}\undb{X}_{p}^{+} (\undb{X}_{p}^{+})^{\dagger} \tilde{\undb{N}}^{\dagger} \undb{U}_{\rm LS} \right]\nonumber \\
  &\stackrel{(b)}{=} \frac{2V_a}{V_p^2 T_p^2} \mathbb{E}\left[\undb{U}_{\rm LS}^{\dagger} \tilde{\undb{N}}\undb{X}_{p}^{\dagger}\undb{X}_{p} \tilde{\undb{N}}^{\dagger} \undb{U}_{\rm LS} \right] \nonumber \\
  &\stackrel{(c)}{=} \frac{2V_a {\rm tr}\left(\undb{X}_{p}^{\dagger}\undb{X}_{p} \right)}{V_p^2 T_p^2} \undb{U}_{\rm LS}^{\dagger} \undb{C}_n  \undb{U}_{\rm LS}  \stackrel{(d)}{=}\frac{2V_a N_t}{V_p T_p} \undb{U}_{\rm LS}^{\dagger} \undb{C}_n  \undb{U}_{\rm LS} \, ,
  \label{kg4}
\end{align}
where we have used $\mathbb{E}\left[\hat{\bm{a}}_A
\hat{\bm{a}}_{A}^{\dagger} \right] = 2V_a \undb{I}_{N_t}$ in
equality $(a)$, and $\undb{X}_p\undb{X}_p^{\dagger} = V_p T_p
\undb{I}_{N_t}$ in equality $(b)$. Furthermore, equality $(c)$
follows from \cite[Lemma 4]{mckay2005general}, and we use ${\rm
tr}\left(\undb{X}_{p}^{\dagger}\undb{X}_{p} \right)= V_p N_tT_p$
in equality $(d)$.

Since the noise covariance matrix $\undb{C}_n $ is unknown, we first find a maximum likelihood (ML) estimate of $\undb{C}_n$, which is then used to estimate $\undb{C}_h$. Given the estimate of the channel matrix $\undb{H}_{\rm LS}$ and the knowledge of the pilot matrix $\undb{X}_p$, the ML estimate of $\undb{C}_n$ is given by
\begin{align}
    \hat{\undb{C}}_n &=  \underset{\undb{C}_n}{{\rm argmax}} \; \, p\left( \undb{Y}_p| \undb{H}_{\rm LS}, \undb{X}_p\right) \nonumber \\
    &= \underset{\undb{C}_n}{{\rm argmax}}\; \prod_{t=1}^{T_p} \frac{\exp\left\{-(\undb{y}^t- \undb{H}_{\rm LS}\undb{x}_p^t )^{\dagger} \undb{C}_{n}^{-1}(\undb{y}^t- \undb{H}_{\rm LS}\undb{x}_p^t ) \right\}}{\pi^{N_r} {\rm det}\left( \undb{C}_{n} \right)} \! .
    \label{ch10}
\end{align}
Taking the log of the likelihood, the equivalent optimization problem is given by
\begin{align}
   \hat{\undb{C}}_n &= \underset{\undb{C}_n}{{\rm argmin}}  \; \Bigg( T_p \log{\rm det}\left(\undb{C}_n\right) + \sum_{t=1}^{T_p} (\undb{y}^t- \undb{H}_{\rm LS}\undb{x}_p^t )^{\dagger} \undb{C}_{n}^{-1}(\undb{y}^t- \undb{H}_{\rm LS}\undb{x}_p^t )  \Bigg) \, .
   \label{ch11}
\end{align}
Taking the matrix variate derivative of the objective function of (\ref{ch11}) with respect to $\undb{C}_n$ and setting it to zero, the ML estimate of $\undb{C}_n$ is given by
\begin{equation}
    \hat{\undb{C}}_n = \frac{1}{T_p} \sum_{t=1}^{T_p} \left(\undb{y}^t-\undb{H}_{\rm LS} \undb{x}_p^t \right)\left(\undb{y}^t-\undb{H}_{\rm LS} \undb{x}_p^t \right)^{\dagger} \, .
    \label{ch12}
\end{equation}
Using the ML estimate of $\undb{C}_n $ in (\ref{kg4}), the estimated value of $\undb{C}_h$ is given by
\begin{equation}
    \hat{\undb{C}}_h = \frac{2V_a N_t}{V_p T_p} \undb{U}_{\rm LS}^{\dagger} \hat{\undb{C}}_n  \undb{U}_{\rm LS} \;,
    \label{kg4a}
\end{equation}
which may be used for estimating
$\sigma_{h,i}^2=0.5\undb{C}_h(i,i)$, as required for estimating
the SKR.


\section{Secret Key Rate Analysis} \label{SKR}
In this section we present the SKR of the MIMO CVQKD system by incorporating the channel estimation errors and the involved overhead. We assume that the entire coherence block is used to generate the secret keys which can then be used for OTP based encryption for data transmission in the subsequent coherence blocks. For generating the secret keys, Alice and Bob begin by generating a correlated random vectors' string $\{\hat{\bm{X}}_{A,n},
\hat{\bm{X}}_{B,n}\}_{n=1}^{N}$ by repeating the quantum key distribution protocol described in section II, $N$ times. 
Given that $T_c$ is the coherence time and $T_p$ is the
pilot overhead, $N$ may be selected as $N=T_c-T_p$. For extracting the final keys, a reconciliation or sifting protocol is carried out by Alice and Bob
over a classical authenticated channel, followed by error correction on the raw keys \cite{weedbrook2010quantum}.
There are two
types of reconciliation protocols: direct reconciliation (DR),
where Alice declares which of the two quadratures should be used
for the secret key generation, and reverse reconciliation (RR),
where Bob declares which of the two quadratures were measured by
him and should be used for the secret key generation on a
classical public channel. It has been previously shown that RR has
a higher SKR than the DR strategy since Eve (who has full control
over the channel) can extract larger information if Alice declares
which of the quadratures should be used for the secret key
\cite{weedbrook2010quantum,ottaviani2020terahertz}. The reason is
that in DR, the signals sent by Alice are accessible to Eve via
the ancilla modes that she injects and are stored in her quantum
memory. However, in RR the measurement outcomes of Bob are not
accessible by Eve. Similar to our initial work
\cite{kundu2021mimo}, here we focus only on RR since positive SKR
that can be achieved by this scheme for any channel transmittance
$\hat{T}_i \in [0,1]$. On the other hand, for DR we require
$\hat{T}_i >0.5$ in order to achieve positive SKRs \cite{weedbrook2010quantum},
which is practically challenging owing to significantly higher path loss (see (\ref{sys_los_nlos})) at THz frequencies \cite{ottaviani2020terahertz}.


In addition to the reconciliation protocol, the SKR also depends
on the type of attack that Eve can perform. The general
assumptions under which the SKRs are evaluated
are\cite{lodewyck2007quantum}: (i) Eve has unlimited computational
power, (ii) Eve has full access to the quantum channel, (iii)
Alice and Bob use an authenticated classical channel for error
correction and information reconciliation, and (iv) Eve cannot
access the apparatuses used by Alice and Bob in their respective
laboratories. There are two types of attacks which Eve can
implement and these are ranked in terms of the increasing amount
of information that Eve can extract. These attacks depend on how
Eve interacts with the individual signals sent by Alice and when
she measures the ancilla mode stored in her quantum memory. Here,
for both types of attacks, we generalize the SKRs of the SISO
system carried out in \cite{lodewyck2007quantum,djordjevic} for
our proposed MIMO system.

\subsection{Eve Attack Mode I: Individual Attack}
Individual attack is the weakest attack which Eve can implement. Here, she individually measures each incoming signal from Alice and the ancilla output is stored in a quantum memory. In order to extract the key information she measures the ancilla mode before the error correction step but after the reconciliation protocol carried out by Alice and Bob. For individual attack, the maximum key information accessible to Eve is given by the Shannon's mutual information between Eve's and Bob's
measurement outcomes. The optimal individual attack is given by
the Gaussian individual attack \cite{lodewyck2007quantum}. When
Eve implements an individual attack, the SKR of the $i$-th
parallel channel (in RR) is expressed as
\begin{align}
    R^{I}_i &= \left(1-\frac{T_p}{T_c}\right) \Big( \beta I\left( X_{A,i}:X_{B,i}\right)- I\left(X_{B,i} :E_i\right) \Big) \, , \quad  i=1,\ldots,r \, ,
    \label{indv1}
\end{align}
where $I\left( X_{A,i}:X_{B,i}\right) $ denotes the Shannon's mutual information of Alice's and Bob's measurement outcomes, $I\left( X_{B,i}:E_i\right) $ denotes the Shannon's mutual information of Eve's and Bob's measurement for the $i$-th parallel channel, and $\beta$ is the reconciliation efficiency. Note that the factor $(1-T_p/T_c)$ arises in
(\ref{indv1}) due to the channel estimation overhead. The Shannon's mutual information of Alice's and Bob's measurement outcomes for the $i$-th parallel channel is thus given as
\begin{equation}
   I\left( X_{A,i}:X_{B,i}\right) = \frac{d}{2} \log_2\left(1+ \frac{\hat{T}_iV_s}{\Lambda_i\left(V_0,W \right)+\sigma_{\rm det}^2 + \sigma_{h,i}^2}  \right) \;,
   \label{indv2}
\end{equation}
where $\Lambda_i(x,y)\dn \hat{T}_ix+(1-\hat{T}_i)y$, and $d$ is
the measurement parameter which takes the value $d=1$ for homodyne
measurement and $d=2$ for heterodyne measurement. Since in the
individual attack Eve measures the ancilla just after Bob reveals
the quadratures measured by him and before the error correction,
the maximum accessible information to Eve is restricted by the
Shannon's information obtained from her ancilla. Eve's information
in RR is given by
\begin{equation}
 I\left(X_{B,i} :E_i\right) = \frac{d}{2}\log_2\left( \frac{V_{B}^i}{V_{B|E}^i} \right)  \, ,
 \label{indv3}
\end{equation}
where $V_{B}^i = \Lambda_i\left(V_a,W \right) + \sigma_{h,i}^2 +
\sigma^2_{\rm det}$ is the variance of the Bob's received string
and $V_{B|E}^i = \frac{1}{ \Lambda_i\left(1/V_a,W \right) +
\sigma_{h,i}^2} + \sigma_{\rm det}^2 $ is the conditional variance
of Bob's received string given Eve's measurement for the
$i$-th parallel channel
\cite{grosshans2004continuous,lodewyck2007quantum}. The overall
SKR of the MIMO QKD system when Eve implements an individual
attack is given by
\begin{align}
  R^{I}_{{\rm MIMO}}& =  \sum_{i=1}^{r} R^{I}_i =\left(1-\frac{T_p}{T_c}\right) \sum_{i=1}^{r} \Bigg( \beta \frac{d}{2} \log_2\left(1+ \frac{\hat{T}_iV_s}{\Lambda_i\left(V_0,W \right)+\sigma_{\rm det}^2 + \sigma_{h,i}^2}  \right) \nonumber \\
  & \quad \quad \quad \quad \quad \quad \quad \quad \quad \quad \quad \quad \quad \quad \quad \quad  - \frac{d}{2} \log_2\left(\frac{\Lambda_i\left(V_a,W \right) + \sigma_{h,i}^2 + \sigma^2_{\rm det}}{\frac{1}{ \Lambda_i\left(1/V_a,W \right) + \sigma_{h,i}^2} + \sigma_{\rm det}^2}  \right) \Bigg) \, .
  \label{indv4}
\end{align}


For a better understanding of the effect of the various important system
parameters on the SKR, we find the first order Taylor Series
expansion of the SKR with individual attack. In the low channel
transmittance limit (i.e., $\hat{T}_i\rightarrow 0$), the SKR can
be approximated as expressed by
\begin{align}
  R^{I}_{{\rm MIMO}}& \! \approx \!   \left(1-\frac{T_p}{T_c}\right)\frac{d}{2 \ln(2)} \sum_{i=1}^{r} \Bigg( \! \left( \! \frac{\beta V_s + W-V_a}{\sigma_{\rm det}^2+\sigma_{h,i}^2+W } + \frac{V_a W-1}{V_a(\sigma_{h,i}^2+W)\big(1+ \sigma_{\rm det}^2(\sigma_{h,i}^2+W)\big)} \right)\hat{T}_i \nonumber \\
  & \quad \quad \quad \quad \quad \quad \quad \quad \quad \quad \quad \quad \quad \quad \quad \quad   - \ln\left(\frac{(\sigma_{\rm det}^2+\sigma_{h,i}^2+W)(\sigma_{h,i}^2+W )}{1+ \sigma_{\rm det}^2(\sigma_{h,i}^2+W)} \right) \Bigg) .
  \label{indv5}
\end{align}
The simplified expression (\ref{indv5}) reveals that
the presence of the additional noise terms due to channel
estimation error $\sigma_{h,i}^2$ and detector noise $\sigma_{\rm
det}^2$ degrades the overall SKR of the system. Further, it
reveals that the SKRs are almost the same for both homodyne
$(d=1)$ and heterodyne detection scheme $(d=2)$, since the
detector noise $\sigma_{\rm det}^2$ increase by a factor of $d$
that balances out the factor of $d$ in the numerator of
(\ref{indv5}). This observation is also confirmed in our
simulation results shown in Section \ref{simulation}.


An asymptotic upper bound on the SKR with individual attack that assumes perfect channel knowledge and no detector noise can be found by setting $\sigma_{h,i}^2= \sigma_{\rm det}^2=0$ in the SKR expression of (\ref{indv4}). This SKR upper bound is given by
\begin{align}
 R^{I, {\rm UB}}_{{\rm MIMO}} &= \sum_{i=1}^{r} \frac{1}{2} \Bigg( \log_2\left(1+ \frac{T_iV_s}{T_iV_0+(1-T_i)W}  \right) \nn \\
 & \qquad \qquad - \log_2\big(  \left( T_iV_a+(1-T_i)W\right)\left(T_i/V_a +(1-T_i)W\right)  \big) \Bigg) \, .
 \label{indv6}
\end{align}
For a rank-$1$ MIMO channel, the SKR expression in (\ref{indv6}) is the same as that of a SISO system derived in \cite[Eq.~6.124]{djordjevic2019physical}. We study the effect of channel estimation error and pilot overhead on the SKR performance by comparing the SKR obtained from (\ref{indv4}) with the upper bound (\ref{indv6}) in the simulation results section.
\subsection{Eve Attack Mode II: Collective Attack}
Collective attack is the next strongest attack implementable by Eve in order to extract the maximum key information. Here Eve individually measures each incoming signal from Alice, but she performs an optimal collective measurement on the
collection of stored ancilla after the key distillation procedure.
For this attack, the maximum key information that Eve can extract is given
by the Holevo's information between Eve's and Bob's states. When
Eve implements a Gaussian collective attack, the SKR of the
$i$-th parallel channel (in RR) is obtained as
\begin{align}
    R^{C}_i &= \left(1-\frac{T_p}{T_c}\right) \Big( \beta I\left( X_{A,i}:X_{B,i}\right)-\chi \left(X_{B,i} :E_i\right) \Big) \, , \quad i=1,\ldots,r \, ,
    \label{rate1}
\end{align}
where $I\left( X_{A,i}:X_{B,i}\right) $ is given in (\ref{indv2}). Further, $\chi\left( X_{B,i}:E_i\right) $ is the Holevo information between Eve and Bob's quantum state for the $i$-th parallel channel, that admits
\begin{equation}
    \chi\left( X_{B,i}:E_i\right) = S \left( E_i \right) - S \left( E_i \big| X_{B,i} \right) \, ,
    \label{rate3}
\end{equation}
where $S \left( E_i \right)$ is the von Neumann entropy of Eve's
state and $S \left( E_i \big| X_{B,i} \right)$ is the von Neumann
entropy of Eve's state given Bob's measurement, which can be
either homodyne or heterodyne.

Let $\hat{\rho}_{E,i}$ and $\hat{\rho}_{AB,i} $ be the density
matrices of Eve's state and Alice-Bob's joint state, respectively,
for the $i$-th parallel channel. Similar to the analysis carried
out in \cite[Sec.8.2]{djordjevic}, we assume that Eve has access
to the purification of Alice-Bob's joint state $\hat{\rho}_{AB,i}
$ such that the density matrix of the resulting state is given by
$\hat{\rho}_{ABE,i} = \ket{\psi} \bra{\psi}$. The density matrix
of Eve's state can be obtained by carrying out the partial trace
with respect to (w.r.t) the Alice-Bob subspace, i.e.,
$\hat{\rho}_{E,i} = {\rm tr}_{AB}\left(\hat{\rho}_{ABE,i} \right)=
{\rm tr}_{AB}\left(\ket{\psi} \bra{\psi} \right)$. Similarly, the
joint Alice-Bob state can be obtained by carrying out the partial
trace w.r.t to Eve's subspace, i.e., $\hat{\rho}_{AB,i} = {\rm
tr}_{E}\left(\hat{\rho}_{ABE,i} \right)={\rm
tr}_{E}\left(\ket{\psi} \bra{\psi} \right) $. Thus, Eve's density
operator $\hat{\rho}_{E,i}$ and Alice-Bob's density matrix
$\hat{\rho}_{AB,i}$ have the same eigenvalues, which implies that
both have the same von Neumann entropy. Thus, in order to
evaluate the von Neumann entropy of Eve's state, it is sufficient to
compute the von Neumann entropy of the Alice-Bob subsystem which does not
depend on the measurement outcome of Bob. Further, the covariance matrix of the
Alice-Bob Gaussian state for the $i$-th correlated string is given
as
\begin{equation}
    \bm{\Sigma}_{AB}^{i} = \begin{bmatrix}
    V_a \undb{I}_2 &  \undb{C}_i \\
     \undb{C}_i^T & b_i\undb{I}_2
    \end{bmatrix} \;,
    \label{rate4}
\end{equation}
where
\begin{equation}
    \undb{C}_i = \sqrt{\hat{T}_i\left(V_a^2-1 \right)}   \begin{bmatrix}
    1 & 0\\
    0 & -1
    \end{bmatrix}
\end{equation}
and
\begin{equation}
    b_i=  \Lambda_i\left(V_a,W \right) + \sigma_{h,i}^2 \;.
\end{equation}
The von Neumann entropy of a Gaussian quantum system can be evaluated by determining the symplectic eigenvalues of
the covariance matrix. The symplectic eigenvalues $\lambda_{1}^i,
\lambda_2^i$ of $\bm{\Sigma}_{AB}^{i}$ can be determined by
evaluating the eigenvalues of the matrix
$|i\bm{\Omega}\bm{\Sigma}_{AB}^{i}|$, where the modulus is in the
operatorial sense \cite{weedbrook2012gaussian}. Here,
$\bm{\Omega}$ is the symplectic matrix that admits
\cite{weedbrook2012gaussian} 
\begin{equation}
\bm{\Omega}=\bigoplus_{k=1}^{2}\begin{bmatrix}
    0 & 1\\
    -1 & 0
    \end{bmatrix}\;,
\end{equation}
where $\bigoplus$ denotes the matrix direct sum operation. For a general covariance matrix of the form
\begin{equation}
    \bm{\Upsilon} = \begin{bmatrix}
    \bm{\alpha} & \bm{\gamma}\\
    \bm{\gamma}^T & \bm{\rho}
    \end{bmatrix}\;,
\end{equation}
the symplectic eigenvalues admit
\begin{equation}
  \nu_{1,2} = \sqrt{\frac{1}{2} \left( \Delta \pm \sqrt{\Delta^2-4 \text{det}\bm{\Upsilon}} \right)}  \;,
\end{equation}
where $\Delta = \text{det}\bm{\alpha} + \text{det}\bm{\rho} + 2\text{det}\bm{\gamma}$ \cite{weedbrook2012gaussian}. Using similar calculation for our case, the symplectic eigenvalues $\lambda_{1}^i, \lambda_2^i$ admit
\begin{equation}
    \lambda_{1,2}^{i} = \sqrt{\frac{1}{2}\left(A^i \pm \sqrt{(A^{i})^2-4B^i} \right)}
     \label{rate7} \;,
\end{equation}
where
\begin{align}
    A^i &= V_a^2\left(1-2 \hat{T}_i\right) + 2\hat{T}_i+\left(\Lambda_i\left(V_a,W \right) +\sigma_{h,i}^2\right)^2 \;, \nonumber \\
    B^i &= \left(\Lambda_i\left(1,V_aW \right) + V_a\sigma_{h,i}^2\right)^2 \;.
    \label{rate6}
\end{align}
Finally, the von Neumann entropy of Eve's state $S \left( E_i \right)$ is given by
\begin{equation}
 S \left( E_i \right) = h\left( \lambda_1^i  \right)   + h\left( \lambda_2^i  \right) \, ,
 \label{rate8}
\end{equation}
where $h(x)$ is the function defined as
\begin{equation}
    h(x) = \frac{(x+1)}{2}\log_2\frac{(x+1)}{2}- \frac{(x-1)}{2}\log_2\frac{(x-1)}{2} \;.
  \label{rate9}
\end{equation}

The von Neumann entropy of Eve's state given Bob's measurement $S
\left( E_i \big| X_{B,i} \right)$ depends on the type of
measurement used by Bob which can be either homodyne or
heterodyne. Since homodyne and heterodyne measurements are rank-1 projections, the conditional state of Alice and
Eve given Bob's measurement outcome $\rho_{AE|X_{B,i}}$ is a pure
state \cite{pirandola2021limits}. Therefore the conditional von
Neumann entropy Eve's state given Bob's measurement is equal to
the conditional von Neumann entropy Alice's state given Bob's
measurement, i.e., $S \left( E_i \big| X_{B,i} \right) = S \left(
A_i \big| X_{B,i} \right) $. Therefore, the the symplectic
eigenvalues of the conditional covariance matrix of Alice's state given Bob's measurement outcome need to be evaluated in order to evaluate $S \left( E_i \big| X_{B,i} \right)$.
Using the analysis from
\cite{weedbrook2012gaussian} for general Gaussian measurements,
Alice's conditional covariance matrix when Bob performs homodyne
measurement is given by
\begin{equation}
    \bm{\Sigma}_{A|X_{B,i}}^{\rm hom} = V_a \undb{I}_2 - (b_i+v_{el})^{-1}\undb{C}_i\bm{\Pi}\undb{C}_i^T \;,
     \label{rate10}
\end{equation}
where $\bm{\Pi} := {\rm diag}\left(1,0\right) $. The symplectic eigenvalue of $\bm{\Sigma}_{A|X_{B,i}}^{\rm hom}$ is given by
\begin{equation}
    \lambda_{{\rm hom}}^{i} = \sqrt{{\rm det}\bm{\Sigma}_{A|X_{B,i}}^{\rm hom}} = \sqrt{V_a^2-\frac{V_a\hat{T}_i(V_{a}^{2}-1)}{b_i+v_{el}}} \;.
     \label{rate11}
\end{equation}
When Bob performs heterodyne measurement, the conditional covariance matrix of Alice is given by
\begin{equation}
    \bm{\Sigma}_{A|X_{B,i}}^{\rm het} = V_a \undb{I}_2 - (b_i+2v_{el}+1)^{-1}\undb{C}_i\undb{C}_i^T  \;,
     \label{rate12}
\end{equation}
which upon simplification gives
\begin{equation}
    \bm{\Sigma}_{A|X_{B,i}}^{\rm het} =  \left(V_a -\frac{\hat{T}_i(V_a^2-1)}{b_i+2v_{el}+1} \right) \undb{I}_2 \;.
     \label{rate13}
\end{equation}
The symplectic eigenvalue of  $\bm{\Sigma}_{A|X_{B,i}}^{\rm het}$ admits
\begin{equation}
     \lambda_{{\rm het}}^{i} = V_a -\frac{\hat{T}_i(V_a^2-1)}{b_i+2v_{el}+1} \;.
      \label{rate14}
\end{equation}
Therefore the conditional von Neumann entropy of Eve's state admits
\begin{equation}
    S \left( E_i \big| X_{B,i} \right) = h\left(\lambda_{{\rm hom/het}}^{i}\right) \;,
     \label{rate15}
\end{equation}
where $h(x)$ is the function defined in (\ref{rate9}) and
$\lambda_{{\rm hom}}^{i}$, $\lambda_{{\rm het}}^{i}$ are given by
(\ref{rate11}) and (\ref{rate14}), respectively. Finally, using
(\ref{rate1}) and (\ref{rate3}), the overall SKR of the MIMO QKD system admits
\begin{align}
  R^{C}_{{\rm MIMO}} =  \sum_{i=1}^{r} R^{C}_i  & = \left( 1-\frac{T_p}{T_c}\right) \sum_{i=1}^{r} \Bigg( \beta \frac{d}{2} \log_2\left(1+ \frac{\hat{T}_iV_s}{\Lambda_i\left(V_0,W \right)+\sigma_{\rm det}^2 + \sigma_{h,i}^2}  \right) \nn \\
  & \qquad \qquad \qquad \qquad \qquad \qquad - h(\lambda_1^i) -h(\lambda_2^i)+h(\lambda_{\rm hom/het}^i) \Bigg) \, .
  \label{rate16}
\end{align}

Similar to the individual attack, we find a Taylor series
expansion of the SKR of collective attack to more explicitly
understand the effect of different system parameters on the SKR. In the low channel transmittance limit (i.e., $\hat{T}_i\rightarrow 0$),
the SKR can be approximated as expressed by
\begin{align}
  R^{C}_{{\rm MIMO}}& \approx   \left(1-\frac{T_p}{T_c}\right)\frac{1}{2 \ln(2)} \sum_{i=1}^{r} \left( \left( \frac{\beta d V_s}{\sigma_{\rm det}^2+\sigma_{h,i}^2+W } + \frac{(V_a^2-1)\ln\left(\frac{V_a+1}{V_a-1} \right)}{ (V_a + W+\sigma_{h,i}^2)} \right. \right. \nn \\
  & - \frac{\sigma_{h,i}^2(WV_a-2W^2+1)\ln\left(\frac{W+\sigma_{h,i}^2+1}{W+\sigma_{h,i}^2-1} \right)}{(W+\sigma_{h,i}^2)(V_a+W+\sigma_{h,i}^2)}
  \left. \left. \ - \frac{d(V_a^2-1)\ln\left(\frac{V_a+1}{V_a-1} \right)}{2(W+\sigma_{h,i}^2+\sigma_{{\rm det}}^2)} \right)\hat{T}_i  - h(W+\sigma_{h,i}^2) \right) .
  \label{rate17}
\end{align}
The simplified expression of the SKR with
collective attack in (\ref{rate17}) reveals that in a practical
MIMO CVQKD system, the SKR decreases due to the noise arising from
channel estimation error $\sigma_{h,i}^2$ and detector noise
$\sigma_{{\rm det}}^2$. Similar to the individual attack case, the
simplified expression in (\ref{rate17}) reveals that the SKRs are
almost the same for both homodyne $(d=1)$ and heterodyne detection
schemes $(d=2)$ since the detector noise $\sigma_{\rm det}^2$
increase by a factor of $d$ that balances out the factor of $d$ in
the numerator of the two terms of (\ref{rate17}) that depends on
$\sigma_{\rm det}^2$. This observation is also confirmed in our
simulation results shown in Section \ref{simulation}. Furthermore,
it is easy to verify that in the limit of perfect channel
estimation ($\sigma_{h,i}^2\rightarrow 0$), no detector noise
$(\sigma_{{\rm det}}^2 \rightarrow 0$), and perfect reconciliation
efficiency $(\beta \rightarrow 1)$, the SKR expression in
(\ref{rate17}) is the same as that of the SKR upperbound presented
in \cite{kundu2021mimo}.

\begin{figure}[htp] 
\centering
\includegraphics[width=0.8\textwidth]{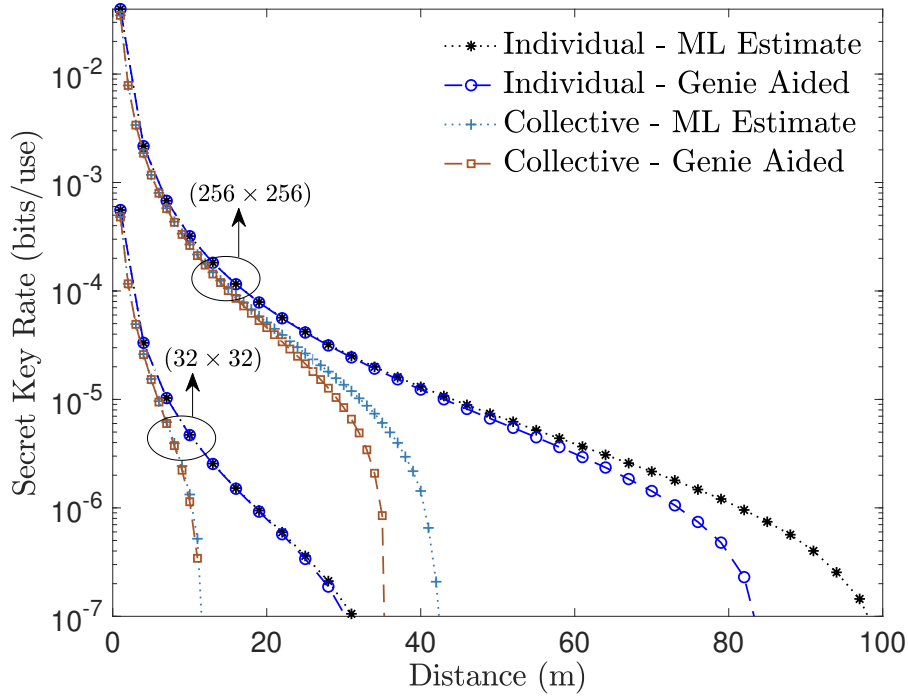}%
\caption{ The plots show the SKR (in bits/channel use) versus distance (m) for two MIMO architectures
$(N_t \times N_r)$ at $f_c=15$ THz. Results are shown for
individual and collective attacks with homodyne measurement using
the ML estimate $\hat{\undb{C}}_n$, and the `Genie Aided' one that
uses true knowledge of $\undb{C}_n$ for evaluating
$\sigma_{h,i}^2$. The other simulation parameters are $V_p=60$ dB,
$W=1$, $T_e=296$ K, $V_s=1$ and $\beta=0.95$
\cite{weedbrook2010quantum}. The antenna gain of each of the
elements at the transmitter and receiver arrays is $G_a=30$ dBi
\cite{rikkinen2020thz,hwu2013terahertz}, $T_p=N_t+500$, and
$T_c=5\times10^5$ \cite{tsujimura2017causal}.
}
\label{fig_genie}
\end{figure}

\begin{figure*}[ht] 
\centering
\subfigure[$32\times32$  ]{%
\includegraphics[width=0.5\textwidth]{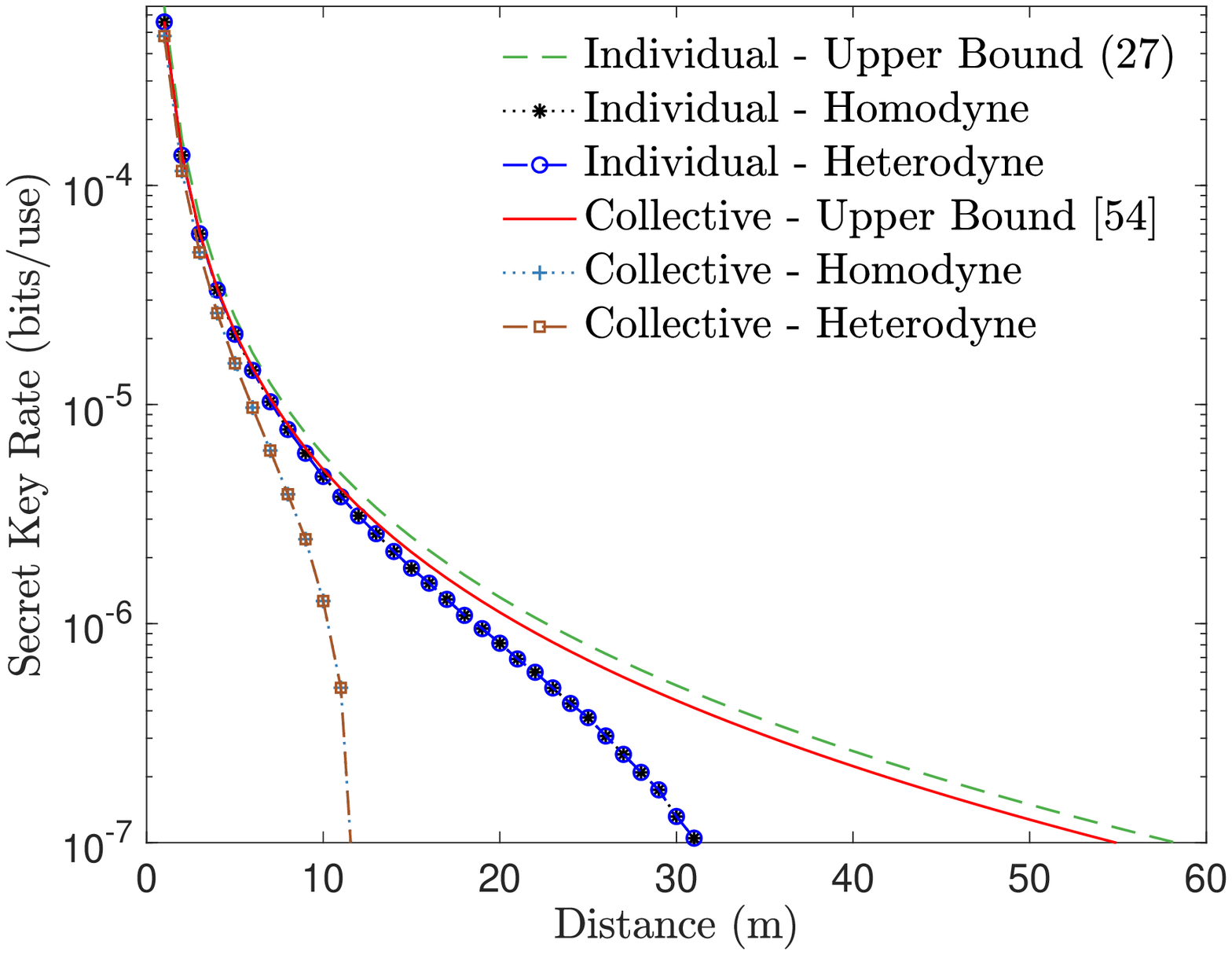}%
\label{fig_sim1:a}%
}\hfil
\subfigure[$256\times256$  ]{%
\includegraphics[width=0.5\textwidth]{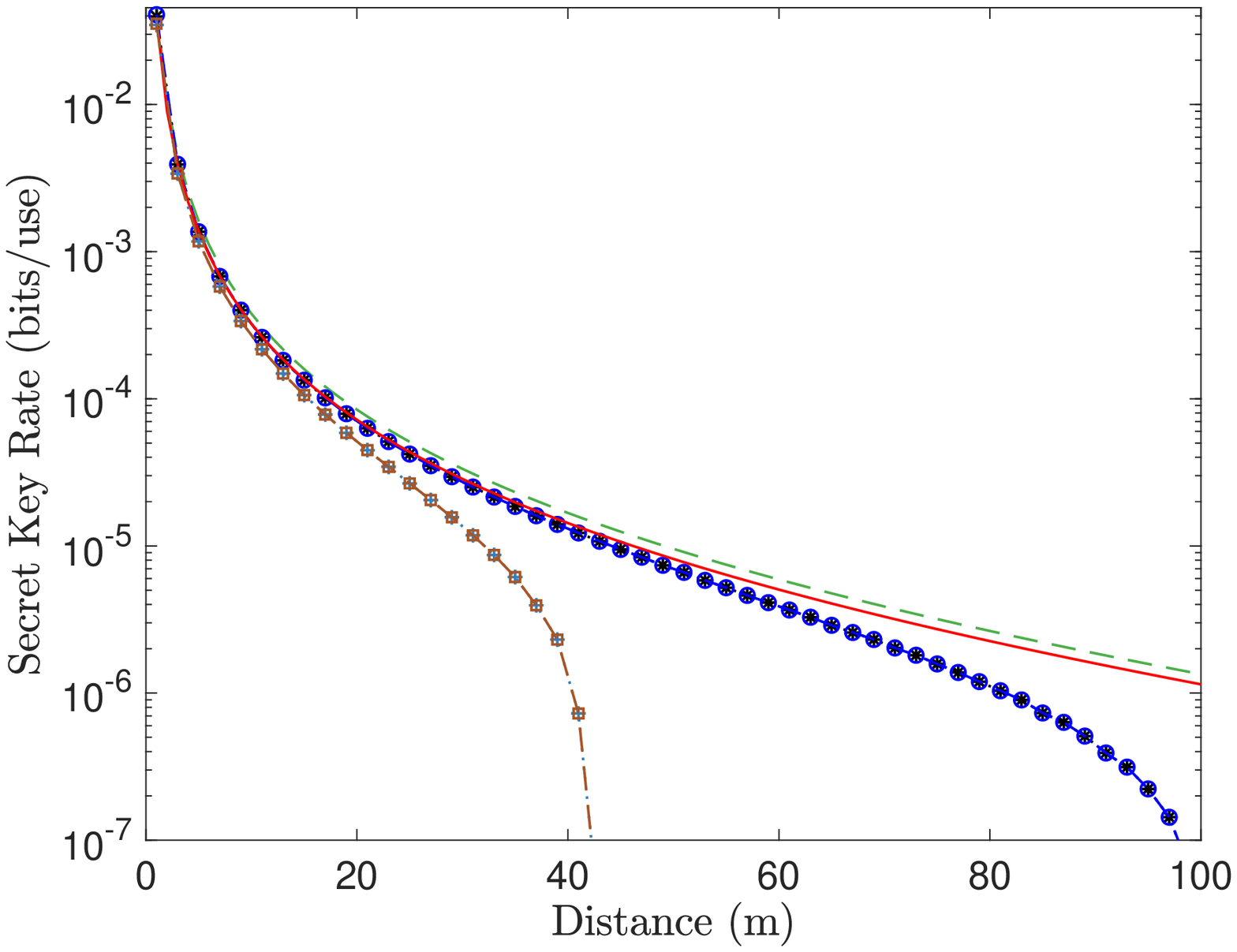}%
\label{fig_sim1:d}%
} \caption{The plots show the SKRs (bits/use) versus
distance (m) for two MIMO architectures $(N_t \times N_r)$
at $f_c=15$ THz. Results are shown for both individual and
collective attacks with homodyne and heterodyne detection. For
comparison we also show the asymptotic SKR upper bound from
\cite[eq.~(20)]{kundu2021mimo} and (\ref{indv6}) for collective
and individual attacks, respectively. The other simulation
parameters are the same as those of Fig. \ref{fig_genie}.
}
\label{fig_sim1}
\end{figure*}

\begin{figure}[ht] 
\centering
\includegraphics[width=0.8\textwidth]{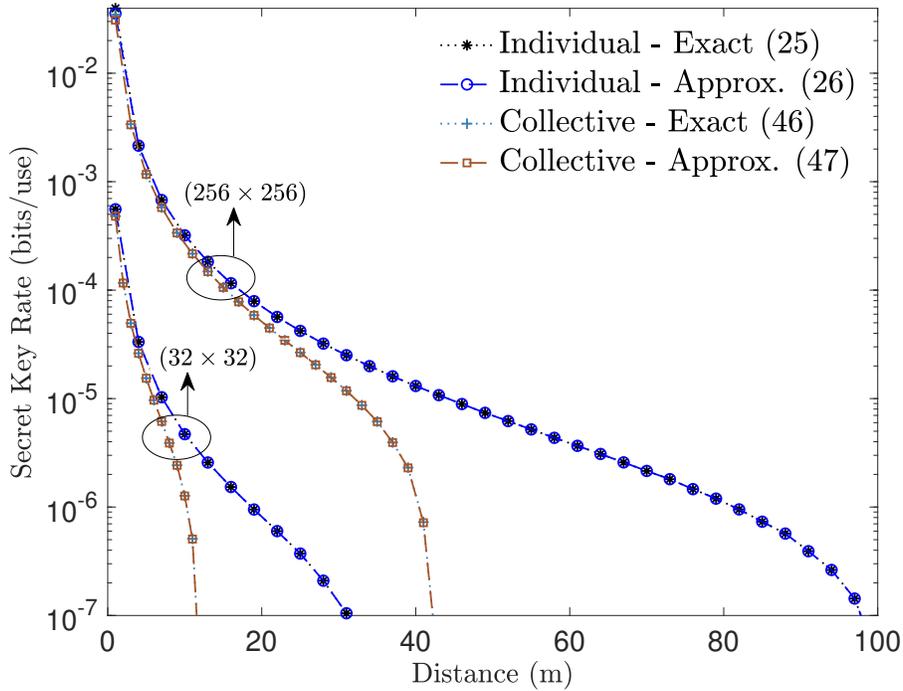}%
\caption{ The plots compare the SKRs (in bits/channel use) versus
distance (m) obtained from the exact and approximate
expressions. Results are shown for both individual and collective
attacks for two different MIMO configurations with homodyne
measurement. The simulation parameters are the same as those of
Fig. \ref{fig_genie}. } \label{fig_expand}
\end{figure}

\begin{figure*}[ht] 
\centering
\subfigure[ Secret Key Rate vs Pilot Duration $T_p$ ]{%
\includegraphics[width=0.5\textwidth]{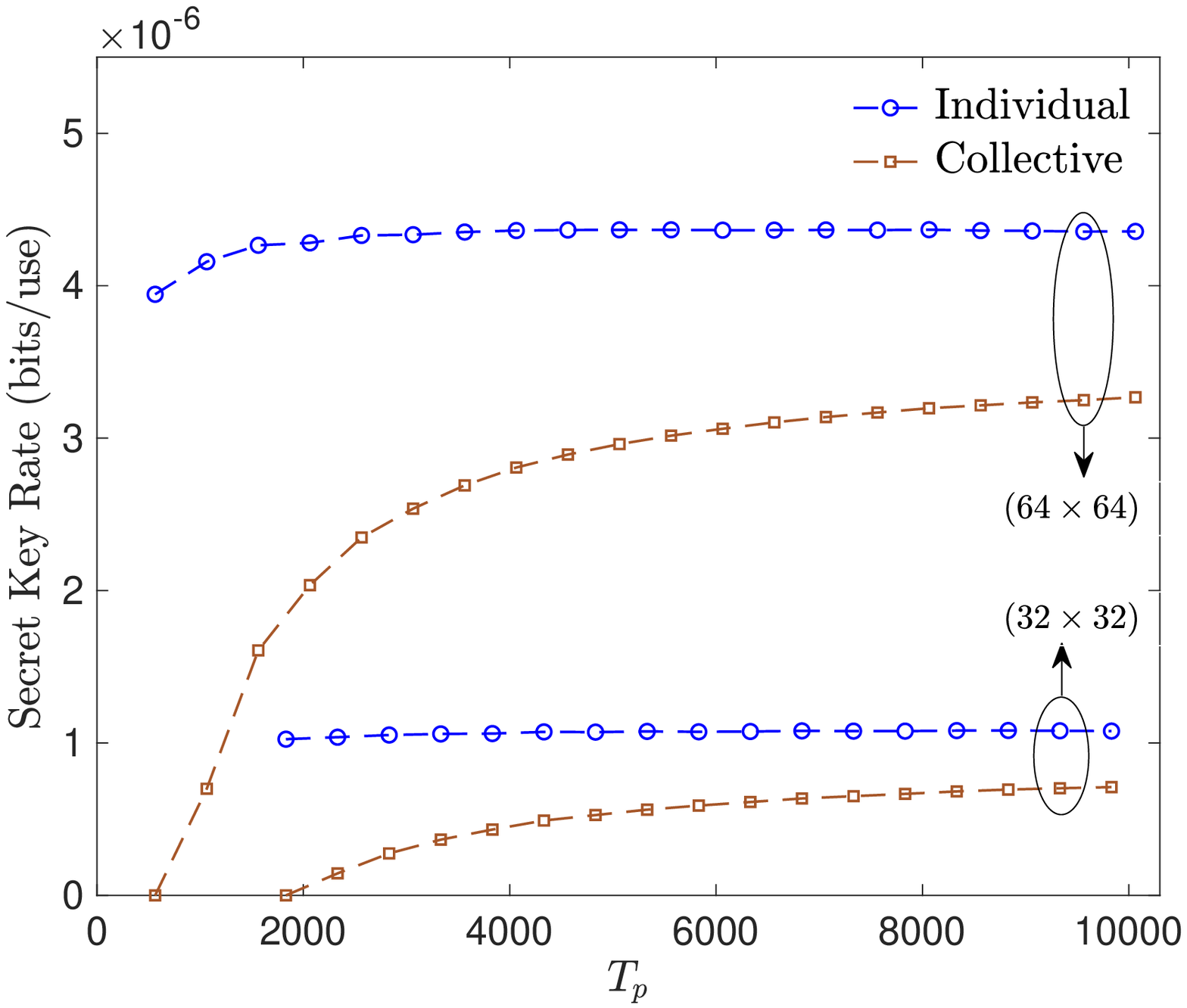}%
\label{fig_sim2:a}%
}\hfil
\subfigure[ Secret Key Rate vs Pilot Power $V_p$ ]{%
\includegraphics[width=0.5\textwidth]{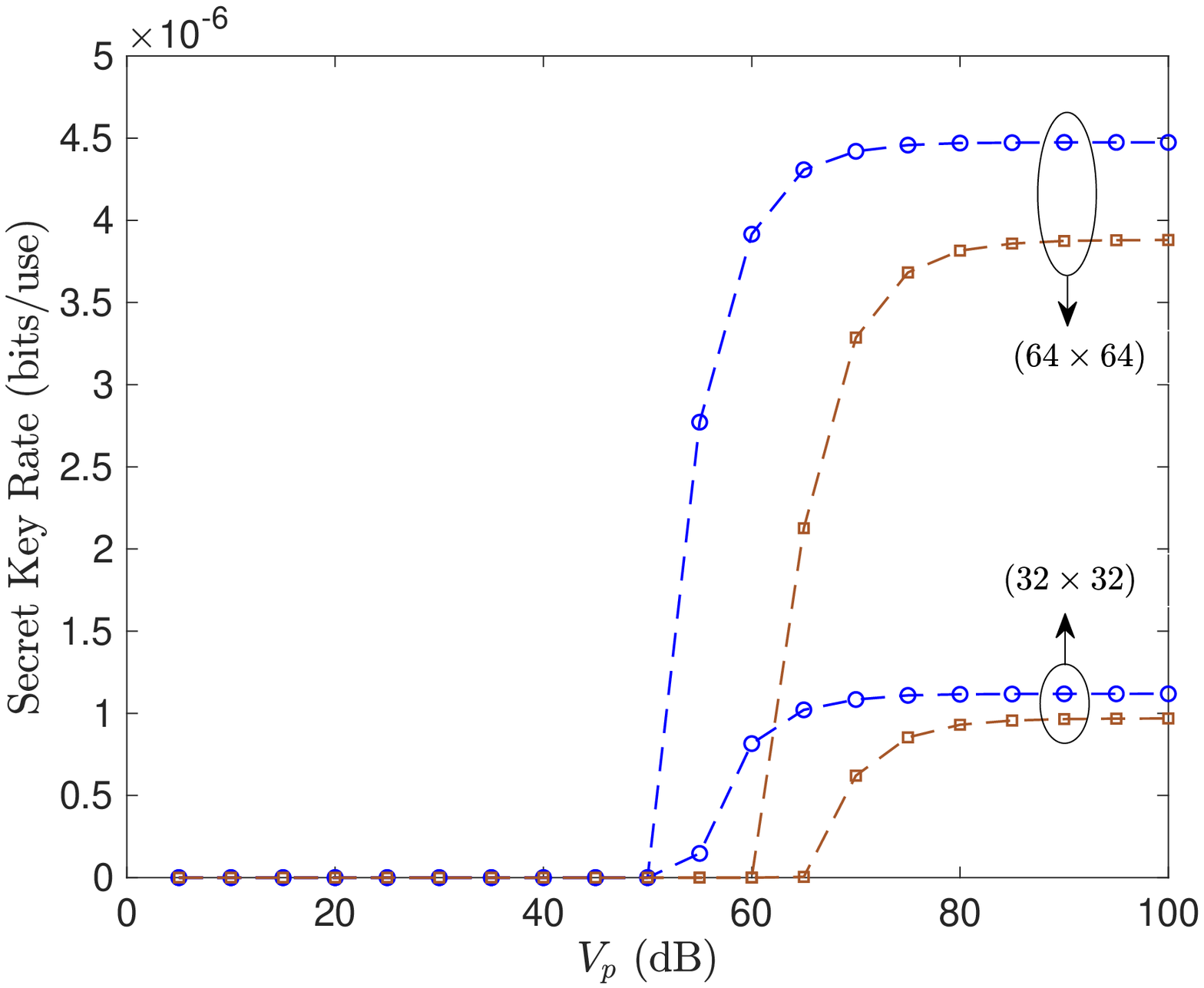}%
\label{fig_sim2:b}%
} \caption{The plots show the SKRs (bits/use) as a function of (a)
pilot duration $T_p$, and (b) pilot power $V_p$ for different MIMO
architectures $(N_t \times N_r)$. Results are shown for a fixed
transmission distance of $d=20$ m at $f_c=15$ THz with homodyne
detection. The rest of the simulation parameters are similar to those used in
Fig. \ref{fig_genie}. } \label{fig_sim2}
\end{figure*}

\section{Simulation Results} \label{simulation}
Similar to \cite{kundu2021mimo}, a simulation scenario with a dominant LoS path is considered with $L=1$. As shown in \cite{kundu2021mimo}, $10-30$ THz is a feasible frequency range that can be utilized to obtain a positive SKR. Here, we show the performance results at $f_c=15$ THz, since the atmospheric absorption coefficient ($\delta=50$ dB/Km) and the free space path loss are lower at $f_c=15$ THz.

We first study the performance of the proposed channel estimation
protocol by plotting the SKR of the MIMO CVQKD system using the ML
estimate of the noise covariance matrix $\hat{\undb{C}}_n$. Fig.
\ref{fig_genie} shows the plot of the SKR (in bits/channel use) versus
distance (m) for two MIMO configurations
at $f_c=15$ THz with homodyne detection. The plots show the SKR
with individual and collective attacks obtained from (\ref{indv4})
and (\ref{rate16}), respectively. The ML estimate uses
$\hat{\undb{C}}_n$ from (\ref{ch12}) in (\ref{kg4a}), and the
`Genie Aided' one uses the true knowledge of $\undb{C}_n$ in
(\ref{kg4}) for evaluating the noise variance due to channel
estimation error $\sigma_{h,i}^2$. It can be observed that at
lower transmission distances, the SKR obtained from the `Genie
Aided' scheme is very close to that of the estimated SKR that uses
the ML estimate $\hat{\undb{C}}_n$. However, at a large
transmission distance for the $(256 \times 256)$ MIMO
architecture, the estimated SKR is slightly higher than the true
SKR (i.e., `Genie Aided'). This is due to the fact that at high
transmission distance the received pilot power is low (due to high
path loss) that leads to a high channel estimation error.
Furthermore, the estimation error of $\hat{\undb{C}}_n$ is high
for the $(256 \times 256)$ MIMO configuration due to the large
dimension of the noise covariance matrix that needs to be
estimated. This estimation error leads to a mismatch between the
true SKR and the estimated SKR, particularly at large transmission
distances. This over-estimate of the SKRs can be mitigated by
increasing the pilot power $V_p$ or the pilot duration $T_p$ at
large transmission distances.

Fig. \ref{fig_sim1} shows the SKR versus transmission distance for
different MIMO configurations and $f_c=15$ THz. It is observed that the
practically achievable SKRs with homodyne and heterodyne
measurements for the two different types of attacks that Eve can
implement. For comparison, we also show the asymptotic SKR upper
bound from \cite[Eq.~(20)]{kundu2021mimo} and (\ref{indv6}) for
collective and individual attacks respectively. It can be observed
that there is a significant gap in the performance of the SKR
upper bound and the practically achievable SKR, particularly at
large transmission distances. This performance gap arises due
additional noise terms due to channel estimation error,
homodyne/heterodyne detector noise, imperfect reconciliation, and
channel estimation overhead. Furthermore, it can be observed that
the performance of the homodyne and heterodyne detection schemes
is almost the same for both individual and collective attacks.
With the heterodyne scheme, the mutual information between Alice
and Bob increases by a factor of two; however, the higher
detection noise compensates this gain and the overall performance
of homodyne and heterodyne schemes are virtually the same. This
observation can also be understood from the approximate SKR
expressions derived in (\ref{indv5}), (\ref{rate17}) for
individual and collective attacks, respectively.

The plots in Fig. \ref{fig_sim1} reveal that although the SKR
upper bound is only slightly better for the individual attack than
the collective attack, the practical SKR performance is
significantly better for the individual attack. Therefore, the
practically achievable SKRs and the maximum transmission distances
can be significantly reduced if Eve has the resources to implement
the stronger Gaussian collective attack.

We now check the accuracy of the approximate SKR expressions
derived in (\ref{indv5}) and (\ref{rate17}) for individual and
collective attacks, respectively. Fig. \ref{fig_expand} shows the
plots of SKR in bits/channel use) versus transmission distance (m)
obtained from the exact (\ref{indv4}), (\ref{rate16}) and
approximate expressions (\ref{indv5}),  (\ref{rate17}) for
individual and collective attacks. Results are shown for two
different MIMO configurations with homodyne measurement. We
observe that the approximate expressions are accurate for
practical transmission distances.

We next study the effect of pilot duration on the SKRs. Since the
simulation results of Fig. \ref{fig_sim1} suggest that for
practical transmission distances at THz frequencies, the SKRs are
very similar for both homodyne detection and heterodyne detection
schemes, here we present only the results for the homodyne case.
Fig. \ref{fig_sim2:a} shows the plot of the SKR for individual and
collective attacks as a function of the pilot duration $T_p$ for
different MIMO configurations at a fixed transmission distance of
$20$ m. From (\ref{kg4}) it can be verified that as $T_p$
increases the noise due to channel estimation error decreases,
which suggests that the SKR should improve as $T_p$ increases. The
simulation results in Fig. \ref{fig_sim2:a} reveal that when Eve
uses an individual attack, the SKR remains almost the same as
$T_p$ increases. On the other hand, the effect of increasing $T_p$
on the SKR is more pronounced for the case of a collective attack
where the SKR first increases as $T_p$ increases and then
saturates to a constant value. Therefore, the effect of channel
estimation error on the SKR is more pronounced for the collective
attack scenario than for the individual attack scenario. In a
practical setting, it is desirable to have a smaller pilot
duration since the computational complexity of channel estimation
in (\ref{ch5}) is $O(T_pN_rN_t)$ with the optimized choice of
$X_p$ in (\ref{ch9}). Therefore, in practice, the pilot duration
should be chosen as the minimum value of $T_p$ at which the SKR
saturates.

We also study the effect of the pilot power $V_p$ on the SKRs.
Fig. \ref{fig_sim2:b} shows the SKR as a function of $V_p$ for
different MIMO architectures at a fixed transmission distance of
$d=20$ m with homodyne detection. We observe that below a
threshold $V_p$ (that depends on the MIMO configuration), the SKR
is zero since the noise variance due to channel estimation error
is too high. In this region, the SKR is limited by the pilot
power. As $V_p$ increases the SKR increases, and then above a
threshold $V_p$ (that again depends on the MIMO configuration),
the SKR saturates. In this regime, the SKR is limited by the
channel gain $\hat{T}_i$ that is constant at fixed $d=20$ m. As
before, we observe that the noise due to channel estimation error
has a more pronounced effect on the SKR with collective attack as
compared to individual attack. Similar to the pilot duration, in
practice, the pilot power should be chosen as the minimum value of
$V_p$ at which the SKR saturates.
\begin{figure*}[ht] 
\centering
\subfigure[ Individual Attack ]{%
\includegraphics[width=0.5\textwidth]{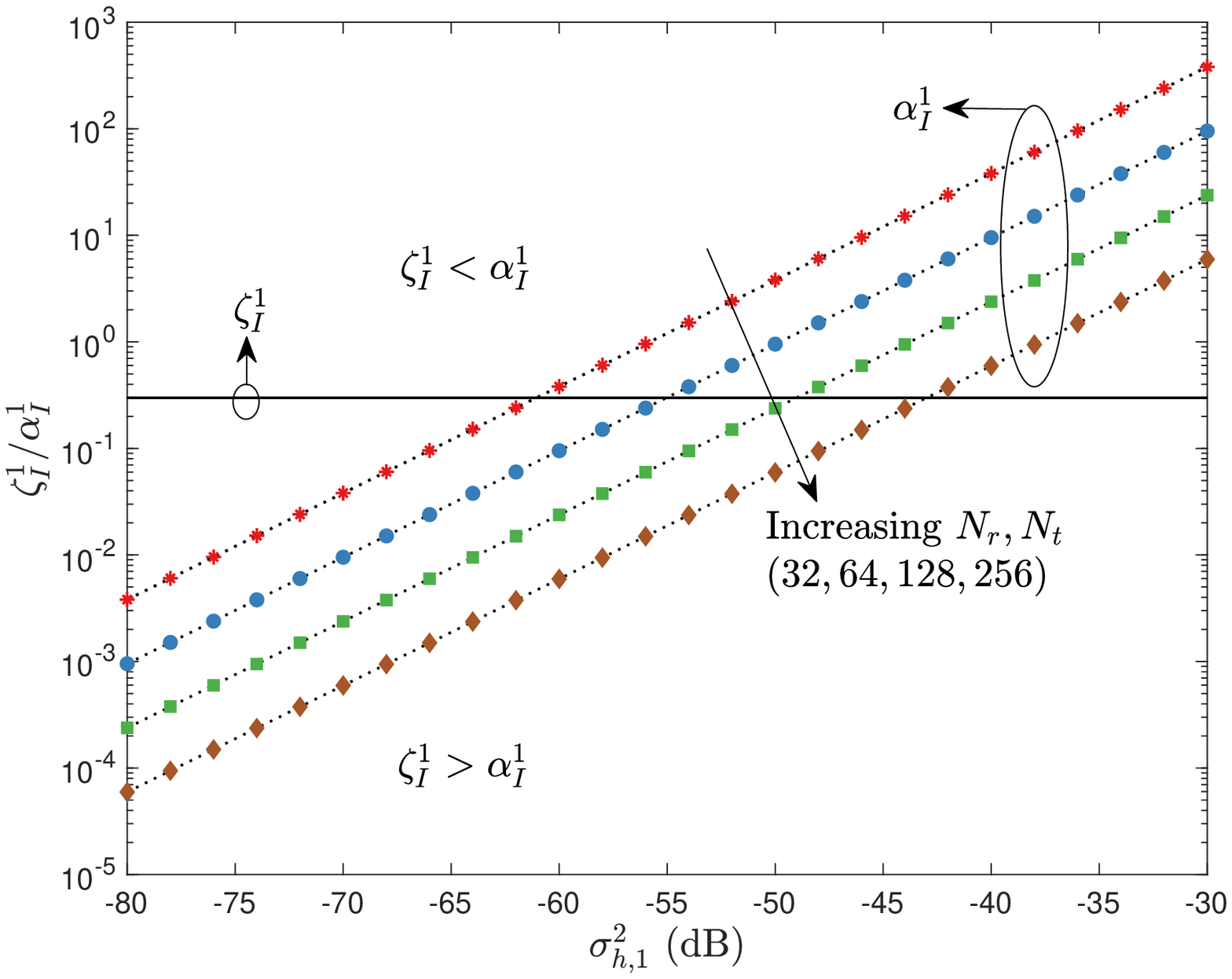}%
\label{fig_sim4:a}%
}\hfil
\subfigure[ Collective Attack ]{%
\includegraphics[width=0.5\textwidth]{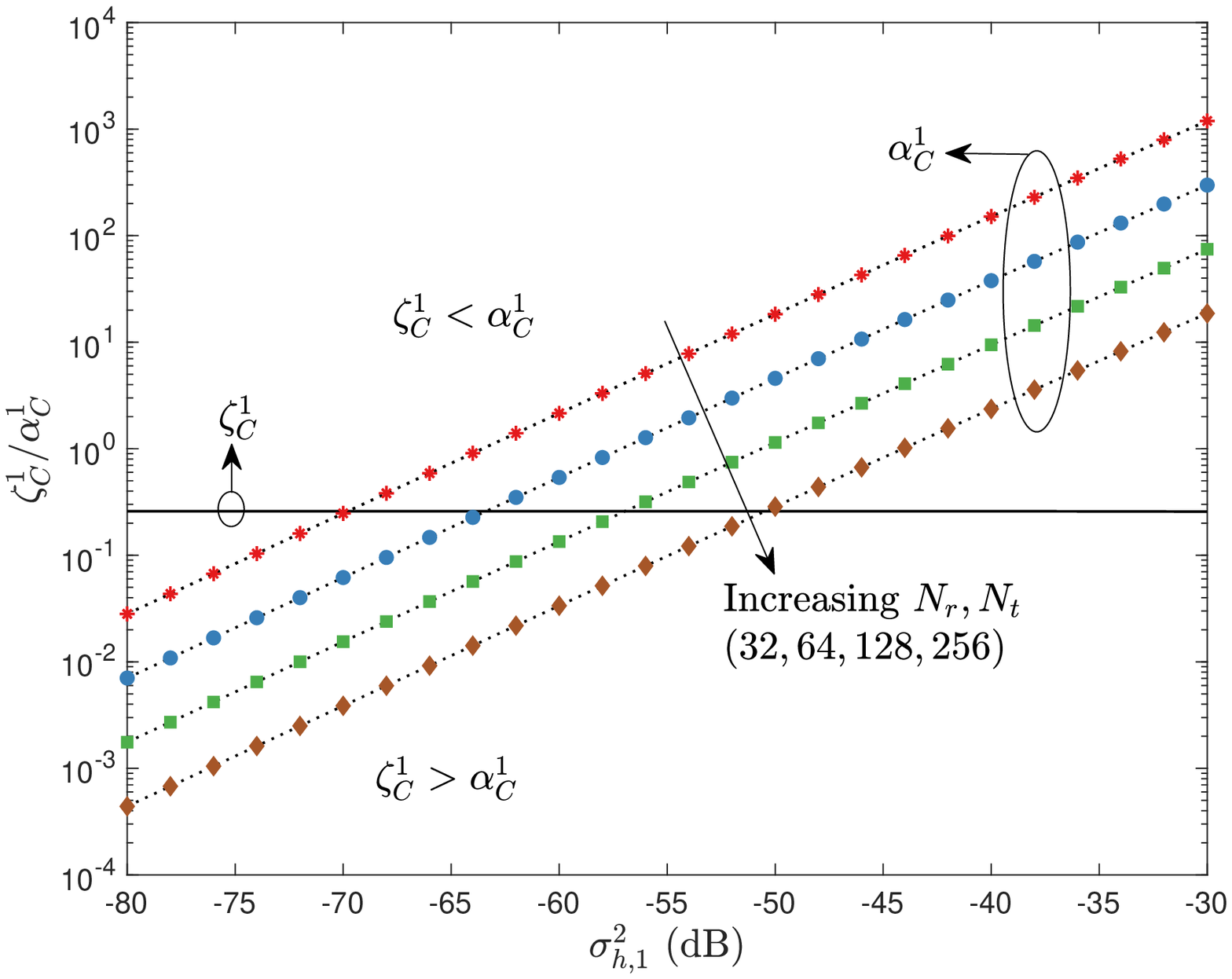}%
\label{fig_sim4:b}%
} \caption{The plots show $\zeta_{C/I}^1$, $\alpha_{C/I}^1$ from
(\ref{simeq1a})-(\ref{simeq2}) versus $\sigma_{h,1}^2$
for various MIMO configurations. Results are shown for both
individual and collective attacks at a fixed transmission distance
of $d=20$ m. Positive SKRs are achievable in the region where the
solid line $\big( \zeta_{I/C}^1\big)$ is above the dashed line
$\big(\alpha_{I/C}^1\big)$. The rest of the simulation parameters are
similar to those used in Fig. \ref{fig_genie}. } \label{fig_sim4}
\end{figure*}

We now numerically study the simplified expressions of the SKRs in
(\ref{indv5}) and (\ref{rate17}) in order to intuitively
understand the effect of the noise from channel estimation error
$\sigma_{h,i}^2$ on the SKRs. We want to find the maximum
tolerable $\sigma_{h,i}^2$ such that positive SKRs can be
achieved. Since the SKRs obtained from homodyne and heterodyne
detection schemes are almost the same, here we consider only
homodyne detection.
Using the simplified SKR expressions from (\ref{indv5}),
(\ref{rate17}), a necessary condition for achieving positive SKR
on the $i$-th parallel channel is given by $ \zeta_{I/C}^i >
\alpha_{I/C}^i $, where $I,C$ denote individual and collective
attacks, respectively. The constants $\zeta_{I/C}^i,
\alpha_{I/C}^i$ admit 
\begin{align}
     \zeta_{I}^i &=   \frac{\beta V_s + W-V_a}{\sigma_{\rm det}^2+\delta_i } + \frac{V_a W-1}{V_a\delta_i\left(1+ \delta_i \sigma_{\rm det}^2\right)} \;,
     \label{simeq1a}
\end{align}
 \begin{align}
    \zeta_{C}^i &=  \frac{\beta V_s -0.5(V_a^2-1)\ln\left(\frac{V_a+1}{V_a-1} \right)}{\sigma_{\rm det}^2+\delta_i } + \frac{ (V_a^2-1)\ln\left(\frac{V_a+1}{V_a-1} \right) }{ (V_a + \delta_i)}     - \frac{\sigma_{h,i}^2(WV_a-2W^2+1)\ln\left(\frac{\delta_i+1}{\delta_i-1} \right)}{\delta_i (V_a + \delta_i)} \;,
    \label{simeq1b}
\end{align}
and
\begin{align}
    \alpha_{I}^i &= \frac{\ln\left(\frac{\delta_i(\sigma_{\rm det}^2+\delta_i)}{1+ \sigma_{\rm det}^2\delta_i} \right)}{\hat{T}_i } \;, \quad
    \alpha_{C} = \frac{h(\delta_i)}{\hat{T}_i} \;,
    \label{simeq2}
\end{align}
where
\begin{equation}
    \delta_i= \sigma_{h,i}^2+W \;.
    \label{simeq3}
\end{equation}

For the simulation scenario considered in Fig. \ref{fig_genie} we
have a rank-$1$ MIMO channel which leads to a single parallel
channel. Therefore, here we study the effect of $\sigma_{h,i}^2$
on $\zeta_{C/I}^i, \zeta_{C/I}^i$ for $i=1$ only. Fig.
\ref{fig_sim4} plots $\zeta_{C/I}^1$, $\alpha_{C/I}^1$ from
(\ref{simeq1a})-(\ref{simeq2}) versus $\sigma_{h,1}^2$
for various MIMO configurations. Results are shown for both
the individual attack and collective attack case at a fixed transmission distance
of $d=20$ m. Here we treat $\sigma_{h,1}^2$ as a free variable
since we want to study the effect of $\sigma_{h,1}^2$ on the SKR
performance. It is easy to verify from
(\ref{simeq1a})-(\ref{simeq3}) that $\zeta_{I/C}^1$ is independent
of the MIMO configuration since it does not depend on $\hat{T}_1$,
whereas $\alpha_{I/C}^1$ does depend on the MIMO configuration.
From Fig. \ref{fig_sim4}, we observe that $\zeta_{I/C}^1$ does not
change much as $\sigma_{h,1}^2$ increases. However,
$\alpha_{I/C}^1$ varies significantly as $\sigma_{h,1}^2$ and the
MIMO configuration changes. The plots in Fig. \ref{fig_sim4}
reveal that positive SKRs are achievable in the region where the
solid line $\big(\zeta_{I/C}^1\big)$ is above the dashed line
$\big(\alpha_{I/C}^1\big)$. We observe that there is a threshold
noise variance $\sigma_{h,1}^2$ above which positive SKRs are not
achievable. Furthermore, we observe that this threshold value of
$\sigma_{h,1}^2$ increases as the number of antennas $N_r, N_t$
increases, since the beamforming gain provided by multiple
antennas increases, which, in turn, increases the magnitude of the
effective channel transmittance $\hat{T}_1$. Hence, the MIMO CVQKD
system can tolerate a much larger $\sigma_{h,1}^2$. Comparing the
plots of Fig. \ref{fig_sim4:a} and Fig. \ref{fig_sim4:b}, we
observe that the threshold $\sigma_{h,1}^2$ is higher for the
individual attack case. Therefore, the MIMO CVQKD system can
tolerate a higher noise variance $\sigma_{h,1}^2$ when Eve
implements an individual attack.

\section{Conclusion} \label{conclusion}
We have proposed a channel estimation protocol for a MIMO THz
CVQKD scheme. The estimated channel matrix is used for SVD-based
transmit-receive beamforming at Alice and Bob. We have
characterized the input-output relation between Alice and Bob by
incorporating the additional noise arising due to channel
estimation error and detector noise. Furthermore, we have analyzed
the SKR of the QKD system under two types of attacks that Eve can
implement: an individual attack and a collective attack. We have
incorporated the finite size effects arising from channel
estimation overhead and imperfect information reconciliation in
the SKR analysis. We have also derived simplified expansions for
the SKRs which are shown to be quite accurate at practical
transmission distances. The simplified expressions are used to
intuitively understand the effect of different system parameters
on the SKR performance of the MIMO CVQKD system. Our simulation
results reveal that the SKR of a practical MIMO CVQKD system
degrades significantly as compared to the asymptotic SKR upper
bound, particularly at large transmission distances. At large
transmission distances, the channel transmittance reduces and the
additional noise variance due to channel estimation error
increases; the combined effect of these two effects degrades the
SKR. Furthermore, our simulation results show that the pilot
duration $T_p$ and pilot power $V_p$ are important system
parameters, since the SKR is zero below a threshold value of
$T_p,V_p$ and the SKR saturates above a threshold value of
$T_p,V_p$. Therefore, the SKR results presented in our paper can
be used to appropriately choose the values of $T_p,V_p$ such that
positive SKRs are achievable in practical THz MIMO CVQKD
implementation.

It is to be noted that we proposed a least-squares based channel
estimation scheme which requires the pilot length to be at least
equal to the number of transmit antennas, i.e., $(T_p\geq N_t)$.
Therefore, the pilot duration overhead can be high for large
dimensional MIMO systems. The pilot overhead can be reduced and
the estimation accuracy can be potentially increased by using
compressive sensing based channel estimation schemes, since the
THz MIMO channel is generally sparse in the angle domain due to a
limited number of scatterers and fewer multi-path components
\cite{sarideentera2021,faisal2020ultramassive}. Therefore, the SKR
analysis of the THz MIMO CVQKD system with compressive sensing
based channel estimation schemes is an important direction to be
studied in future extensions of this work.


\bibliographystyle{IEEEtran}
\bibliography{IEEEabrv,MIMO_QKD}

\end{document}